\documentclass[a4paper,11pt]{article}
\pdfoutput=1 

\usepackage{jinstpub} 
\usepackage[american]{circuitikz} 
\usepackage[mediumqspace,amssymb,pstricks,textstyle]{SIunits}
\title{The MIMA project. Design, construction and performances of a compact hodoscope for muon radiography applications in the context of Archaeology and geophysical prospections.}

\author[a,b]{G.~Baccani}
\author[b,a]{, L.~Bonechi}
\author[b]{, R.~Ciaranfi}
\author[c,d]{, L.~Cimmino}
\author[a,b]{, V.~Ciulli}
\author[a,b]{, R.~D'Alessandro}
\author[b]{, B.~Melon}
\author[c,d]{, P.~Noli}
\author[c,d]{, G.~Saracino}
\author[b]{and L.~Viliani}

\affiliation[a]{Universit\`a di Firenze, Dipartimento di Fisica e Astronomia, \\Via G. Sansone 1, I-50019 Sesto Fiorentino (Firenze), Italy}
\affiliation[b]{Istituto Nazionale di Fisica Nucleare, Sezione di Firenze, \\Via B. Rossi 3, I-50019 Sesto Fiorentino (Firenze), Italy}
\affiliation[c]{Universit\`a degli Studi di Napoli Federico II - Dipartimento di Fisica, Complesso Universitario di Monte Sant'Angelo - Via Cintia, 21 - 80126 - Napoli, Italy}
\affiliation[d]{Istituto Nazionale di Fisica Nucleare, Sezione di Napoli, \\Complesso Universitario di M. S. Angelo, Ed. 6 - Via Cintia, 80126 Napoli, Italy}

\emailAdd{lorenzo.bonechi@fi.infn.it}

\abstract{The Muon Imaging for Mining and Archaeology (MIMA) project aims at the development of a non-invasive technique for imaging dense structures or cavities, hidden in the underground or anyway surrounded by huge volumes of matter, based on Muon Absorption Radiography. Given its natural multidisciplinary, the final purpose is the validation of this methodology for applications in different fields, like Archaeology, Geology, mining, Civil Engineering and Civil Protection, in close cooperation with team in these fields.

In this paper we report on the design, construction and performance of a compact and lightweight muon telescope designed mainly for archaeological investigation and geophysical prospections in general. The MIMA detector is also used currently as a test instrument to study different hardware solutions to optimize the global performance in these types of applications.}


\begin{document}
\maketitle
\flushbottom

\section{Introduction}
\label{sec:intro}
Muon radiography, sometimes referred to as {\it muography} using a term minted by Tanaka \cite{Tanaka}, is conceptually similar to standard X-ray radiography. The role of X-rays is taken here by muons, elementary particles similar to electrons but with a mass about 200 times larger, which are continuously produced in the upper layers of the Earth's atmosphere, mainly in the decays of parent particles like pions and kaons that are generated in the interactions of primary cosmic rays with the Nitrogen or Oxygen nuclei. Being extremely penetrating particles, muons are able to reach the Earth's surface quite easily and are finally absorbed in varying degrees as they pass through hundreds of metres of soil and rock. By measuring the degree of absorption of this natural radiation downstream a target volume, the two dimensional density profile of the traversed matter can be reconstructed, thus allowing also to identify the position of regions with density values significantly different from those expected.

The first use of cosmic ray muons for studying thick layers of material dates back to 1955, when the British phisicist E. P. George measured muon attenuation to determine the overburden of rock above a tunnel inside a mountain \cite{George}. This measurement was followed in 1970's by the firt real muon radiography by the Particle Physics Nobelist Luis Alvarez \cite{Alvarez} in the search for a hidden chamber inside the Chephren pyramid in Egypt. He used a 4\,m$^2$ spark chamber as muon detector, placed at the bottom of the pyramid.

Thanks to the great progress in detection techniques and to the pioneering application to volcanic edifices in Japan \cite{Tanaka}, followed by our group in Italy \cite{muray1,muray2,MURAVES} and other groups in France \cite{cristina} among others, muography has now reached maturity, also being applied nowaday to surveying of many kind of natural or man-made structures. Indeed in 2015 a muographic image of the nuclear reactor at the Fukushima Dai-ichi nuclear power plant, damaged by the 2011 Great Tohoku Earthquake and Tsunami, proved that the nuclear core of the reactor was actually melted during the accident \cite{fukus}. The recent works published by Saracino et al. \cite{echia} and Morishima et al.\cite{kufu} are a representative proof of the maturity of the technique.                         

Muon radiography proceeds like standard medical imaging by measuring the attenuation of a particle beam, represented in this case by the natural flux of atmospheric cosmic ray muons which crosses a material volume (e.g. the dome of a volcano) with a sensitive device or particle detector, like a muon telescope. The measurement of the muon attenuation allows an estimate of the opacity of the structure, which is related to the integral of the density along the muon trajectory in the matter. This is achieved by comparing the muon flux measured after crossing the target, to the incident open sky flux. The presence of materials along the muon trajectory acts as a filter since only the most energetic muons will be able to escape from the structure. 

The development of the MIMA project is meant to address many important aspects that are relevant for the development and application of a highly innovative technique for the non-invasive search of hidden structures, heavy materials or empty cavities in Archaeology, Geology and other fields, with a multidisciplinary approach. For this reason the detector is currently being used, and it will be in the next few years, for testing the proposed methodology in different types of studies. All of these activities are carried on in close cooperation with groups of geologists and archaeologists, with the aim of investigating the differences in terms of performances of muon radiography and the more classical techniques currently used for these kind of studies.


Basically two kinds of detectors are currently used in volcanoes muography: electronic detectors, like the Mu-Ray (e.g. \cite{muray1}) hodoscope, based on technologies developed in the field of high energy particle physics, and nuclear photographic emulsions \cite{Tanaka}. For the MIMA project we have opted to deploy a scaled down replica of the Mu-Ray telescope, with some peculiar choices for improving the performance and allow the installation inside mines, tunnels etc.

\section{Concept and design of the MIMA muon telescope}
\label{sec:concept}

The new generation of muon telescopes designed for muon radiography in inhospitable environments are often characterized by a modular structure with at least three detection modules, each providing position measurements along two orthogonal X and Y axes. The modules are usually mounted on a dedicated platform which allows modifying the measuring direction of the telescope to acquire free-sky muon flux calibration data\cite{muray1, muray2}. In particular a muon hodoscope designed for archaeological and geophysical applications must satisfy several important requirements such as: 

\begin{itemize}
\setlength\itemsep{0em}
\item low energy consumption;
\item robustness;
\item compact and lightweight mechanics for easy transportation;
\item small enough to fit inside boreholes or other particular environments requiring these feature;
\item wide range of operating temperature, from few degrees below zero to 50$\div$60\,$\degree$C;
\item tracking capability suitable to determine the direction of muons with few milliradians angular resolution;
\item background suppression capability (to exclude fake tracks or scattered muons traveling in the wrong versus).
\end{itemize}

MIMA has been designed and built to fulfill all of these requirements. The MIMA project is framed in the context of the imaging of unexcavated sectors of archaeological sites, of the interior of structures or buildings, of the overburden material above galleries or mines, aiming at the search for hidden cavities or volumes containing high density materials. 

The construction of the MIMA telescope has its roots in the expertise gained within activities performed by us for volcanological studies, i.e. the INFN MURAY project and the MURAVES "Progetto Premiale" \cite{MURAVES}. From these activities MIMA has inherited the structure of the detector's tracking plane and the operating method. Currently the powering and DAQ system is derived from that developed for the MURAVES project, but an activity is on-going to simplify all the electronics and further reduce its power consumption, that is anyway already very low, amounting to approximately 30\,W for the whole system. 

The MIMA hodoscope is a compact and lightweight muon telescope, its dimensions are 50\,cm$\times$50\,cm$\times$50\,cm for a total mass of approximately 60\,kg including the mechanics and the electronics. Although two measured points along the muon trajectory could be sufficient for the reconstruction of a track, the MIMA tracker consists of three X-Y tracking modules fixed inside a thin aluminum cube forming a particle tracking telescope with a geometrical factor of the order of 1000\,cm$^2$sr$^{-1}$. The presence of the third (central) X-Y module is important for a clean reconstruction of the events. By requiring a tight alignment of the reconstructed hit coordinates, the presence of the third module allows in fact identifying the eventual multiple tracks appearing in the same event, due for example to the passage of different muons belonging to the same cosmic ray shower, or rejecting fake events that could be also triggered by random coincidences of different particles but also by noise signals (see section~\ref{sec:trigger}).

\begin{figure}[thb]
\centering
\includegraphics[width=0.4\linewidth]{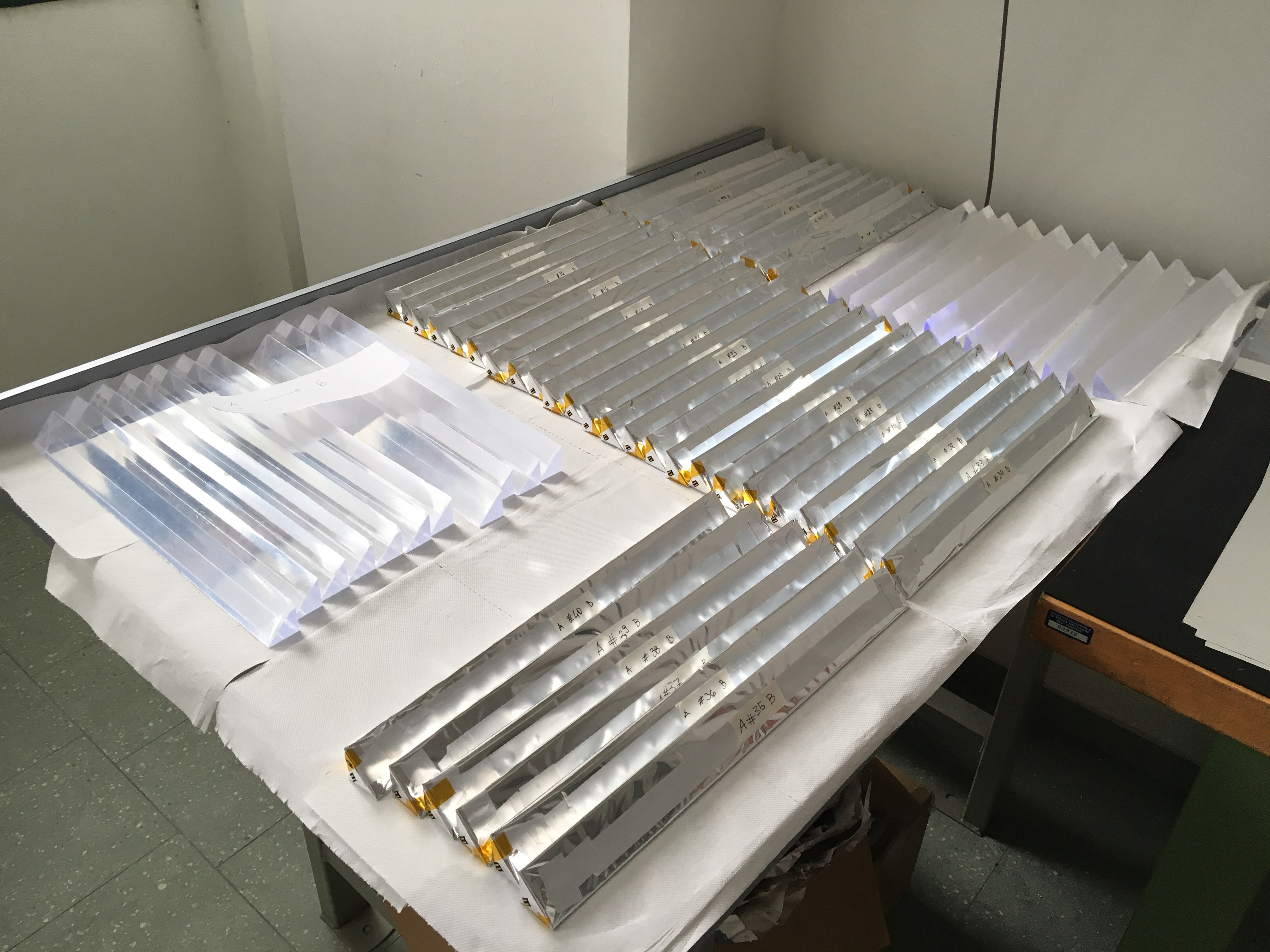}
\qquad
\includegraphics[width=0.4\linewidth]{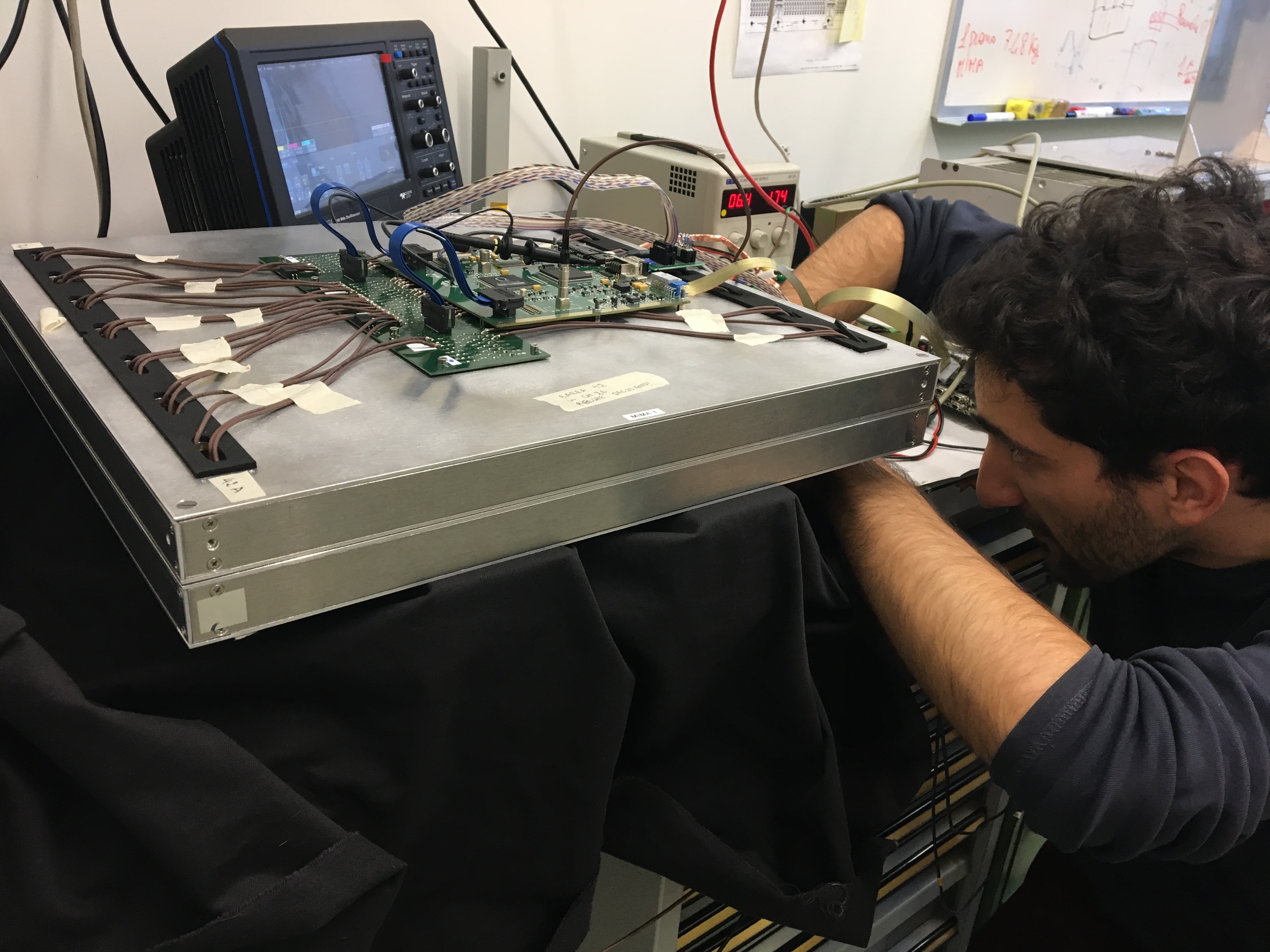}
\caption{Left: the total set of scintillator bars with triangular section used for the MIMA tracking planes. Right: one module composed of X and Y planes under test.}
\label{fig:mima00}
\end{figure}

The basic detection plane consists of 21 plastic scintillator bars assembled in such a way to produce a 40\,$\times$\,40\,cm$^2$ active detection surface. A total of 126 plastic scintillator bars have been used to build the three X-Y modules of MIMA (see left photo of figure \ref{fig:mima00}). Each module is made up of two identical and independent planes, one of which overlaps the other upside down, staying in close contact and with bars rotated of 90$\degree$ (see right photo of figure \ref{fig:mima00}). This configuration defines two X-Y orthogonal axes in order to measure two independent coordinates of the muon impact points (hits).

\begin{figure}[thb]
\centering
\includegraphics[height=0.5\linewidth]{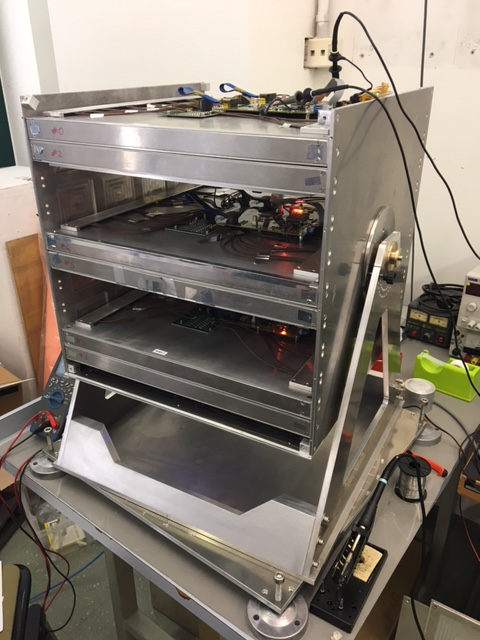}
\qquad
\includegraphics[height=0.5\linewidth]{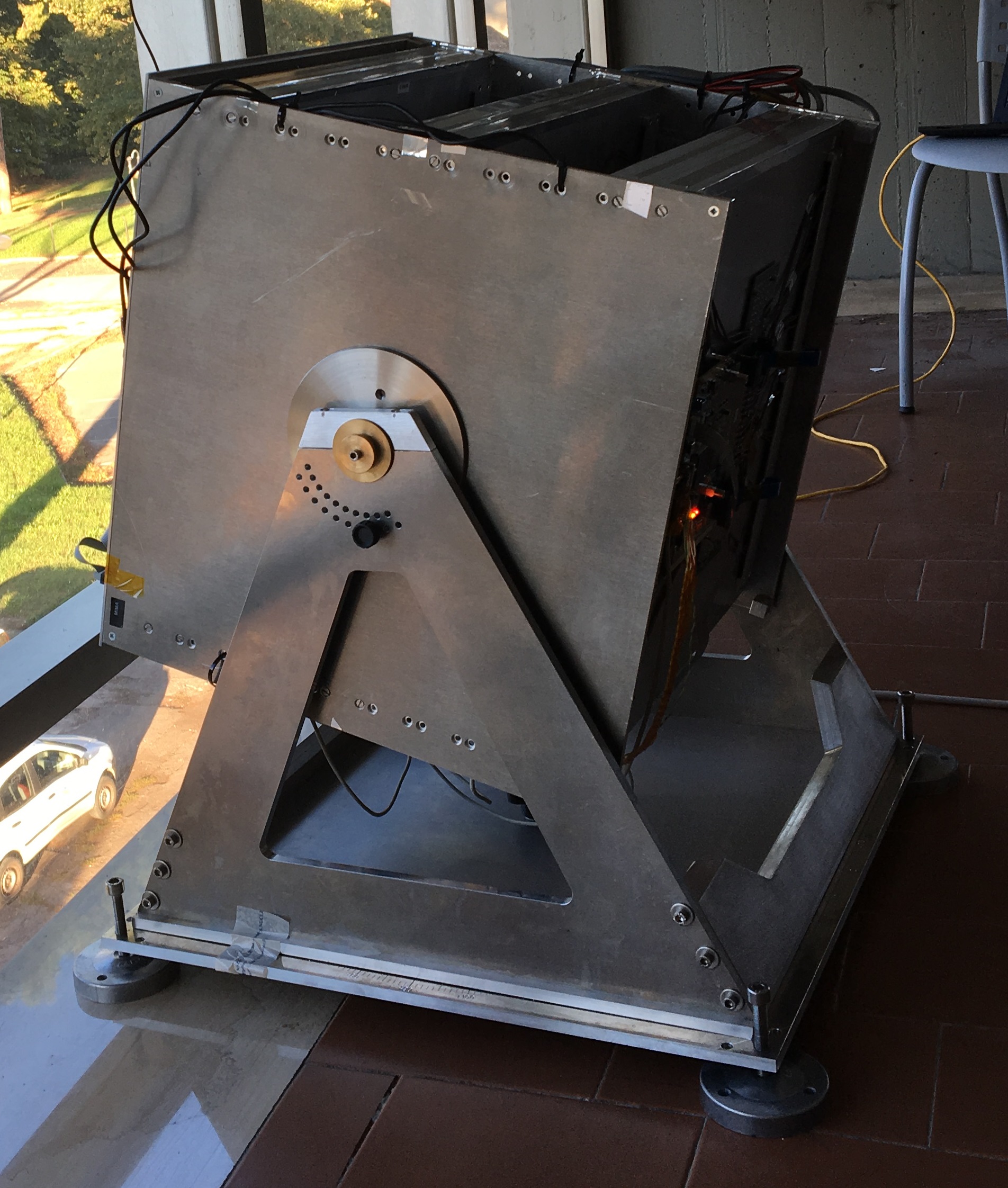}
\caption{Left: the MIMA tracker under test after assembling three complete modules inside the mechanics. Right: the final MIMA telescope, provided of an altazimutal mounting, used for on-field measurements.}
\label{fig:mima01}
\end{figure}

The three MIMA X-Y modules are stacked inside a cubic aluminum frame (left photo of figure \ref{fig:mima01}) mounted on an altazimutal platform  which allows modifying the azimuth angle continuously and the zenith angle with 5$\degree$ steps (right photo of figure \ref{fig:mima01}). In this way it is possible to adapt the pointing direction of the telescope depending on the position of the volume under investigation. Details on the mechanical structure can be found in section \ref{sec:mechanics}.

\subsection{The tracking plane}
\label{sec:tracking_plane}
The MIMA telescope has a modular structure. Each module consists of two independent scintillator planes each of which assembled using polystyrene-based plastic scintillator bars. The choice of a plastic scintillator detector for muon radiography has different reasons. This type of detector does not provide the best spatial and angular resolutions for track reconstruction, but it can be designed to be very robust, suitable to be frequently transported to inhospitable installation locations, and it requires only a minimal maintenance by operators. It can be further readout using low power electronics and it is relatively insensitive to ambient conditions.

\subsubsection{Geometrical configuration}
\label{sec:tracking_plane:geom}
The basic element of the plane is a 40\,cm long plastic scintillator bar with an emission spectrum peaked at around 418\,nm (in the region of violet) and a typical emission time constant of $\simeq$ 3\,ns. Each bar is readout using two silicon photomultipliers (SiPM) photosensors (see sections \ref{sec:sipm} for more details). Differently from the Mu-Ray and MURAVES detectors, thanks to the choice of a very good quality plastic scintillator and to the improvement in the quality of the SiPM devices, in case of MIMA we optically adapted the SiPM sensors directly to the scintillators, thus avoiding the installation of intermediate wavelength shifter (WLS) fibers and obtaining a simpler and cheaper system. Two different kinds of scintillator bar have been used to build the tracker: scintillator bars of triangular shape (with 4.0\,cm basis and 2.0\,cm height), whose idea has been inherited from the previous projects mentioned above, have been used to produce the outer four of the six planes while for the remaining planes, installed in the central part of the tracker, bars of rectangular shape (with 3.0\,cm basis and 1.0\,cm height) have been chosen to test a different original configuration.

\begin{figure}[thbp]
\centering 
\includegraphics[height=.2\textwidth]{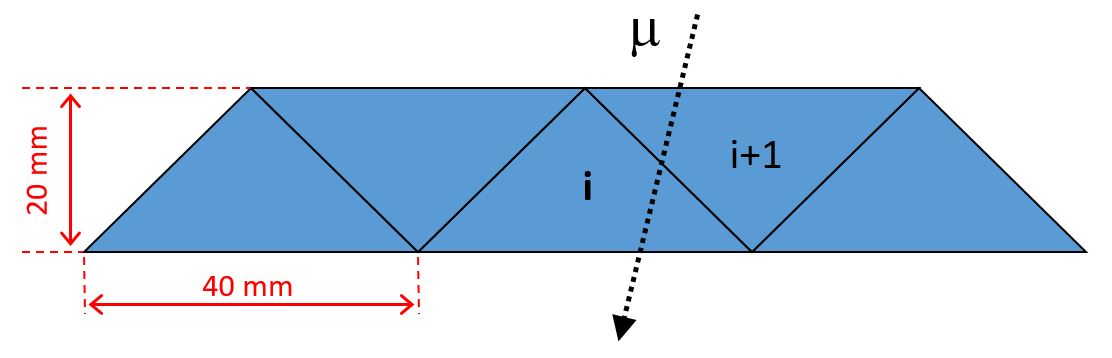}\\
\includegraphics[height=.2\textwidth]{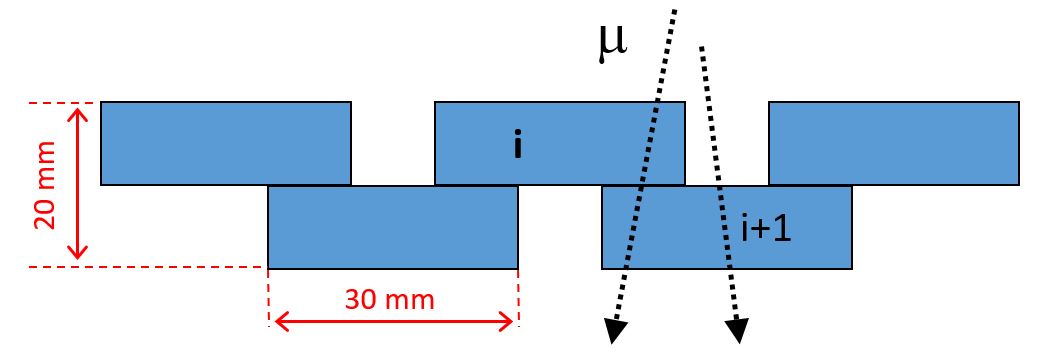}
\caption{Top: geometrical configuration of the scintillating bars with triangular section  for the tracking planes of the outer two modules. Bottom: geometrical configuration of the scintillating bars with rectangular section for the tracking planes of the central module.}
\label{fig:scint_scheme} 
\end{figure}

Concerning the outer four tracking planes, scintillators with triangular cross-section are assembled in such a way to produce an homogeneous 40$\times$40$\times$2\,cm$^3$ fiducial region (top sketch in figure \ref{fig:scint_scheme}) and their spatial configuration allow achieving a spatial resolution of the order of 3\,mm, approximately half of that expected for a digital algorithm considering the 2\,cm spacing between the bars. More details are given in paragraph~\ref{sec:res}. This is obtained by exploiting the weighted average of the signals produced in the front-end electronics by the different amounts of light collected inside two adjacent bars. For the central planes, rectangular bars are assembled instead using a new peculiar configuration, still under test (bottom sketch in figure \ref{fig:scint_scheme}). These bars are positioned in such a way to define three different regions for each bar. Considering the i-th bar and muons traveling orthogonally to the surface of the plane, two types of events can in fact be identified: those hitting only a single bar and b) those hitting two adjacent bars. For a single bar three such independent sectors can be identified in such a way that it is then possible to use a digital algorithm with a spacing parameter which is one third of the bar width (1\,cm in our case), thus smaller than the readout pitch (2\,cm in our case), obtaining finally an RMS value of approximately 3\,mm on the determination of the impact point for vertical tracks. Clearly the situation is more complex for inclined tracks and a dedicated algorithm, which takes into account the different paths inside each bar for different inclinations of the tracks, has been developed to exploit the information on the signal amount released in adjacent bars to improve the spatial resolution. An uncertainty of the order of 3$\div$4\,mm is found for these planes. The uncertainty distribution is not Gaussian and discretization effects are found which depend of the inclination of the tracks.

\subsubsection{Production}
\label{sec:tracking_plane:prod}

While the scintillator bars with rectangular section have been used as coming from the factory, with the purpose of reducing costs, those with triangular section have been cut from 80\,cm long bars with 4\,cm$\times$4\,cm square section, and then polished carefully to have a good optical quality thus improving their light collection capability. This work has been carried on by the mechanics atelier of the Department of Physics and Astronomy of the University of Firenze.
Two silicon photomultipliers are glued at the two far ends of each bar to convert fractions of the scintillating light collected on the smaller faces of the bar by internal reflections into electrical signals. The surfaces of the scintillator bars have been preliminary cleaned from the dust in order to avoid the loss of efficiency since dust may act as trap for the photons. A thin reflective aluminized Mylar film has been used to wrap each single bar to guarantee optical insulation from the environment and from the other bars and to increase at the same time the internal light collection. Very small black plastic frames produces by a 3D-printer have been glued on both the small surfaces of the bars to better define the position of the SiPMs, facilitate their gluing and avoid unexpected contacts with the metalized film, that could provoke dangerous short circuits on the readout electronics (left photo in figure \ref{fig:assembling}). To obtain a good optical coupling between the photodetector and the scintillator surface a transparent silicon sealing compound has been used with a refractive index of 1.49 for 430\,nm wavelength light, similar to that of the scintillator material and of the thin protective layer of epoxy resin covering the active surface of the SiPM). A black silicon adhesive has been deposited between the SiPM edges and the surrounding flaps of the Mylar film to better fix all parts, thus improving also the optical insulation.

\subsection{Mechanical structure} 
\label{sec:mechanics}

\begin{figure}[th]
\centering 
\includegraphics[width=.43\textwidth]{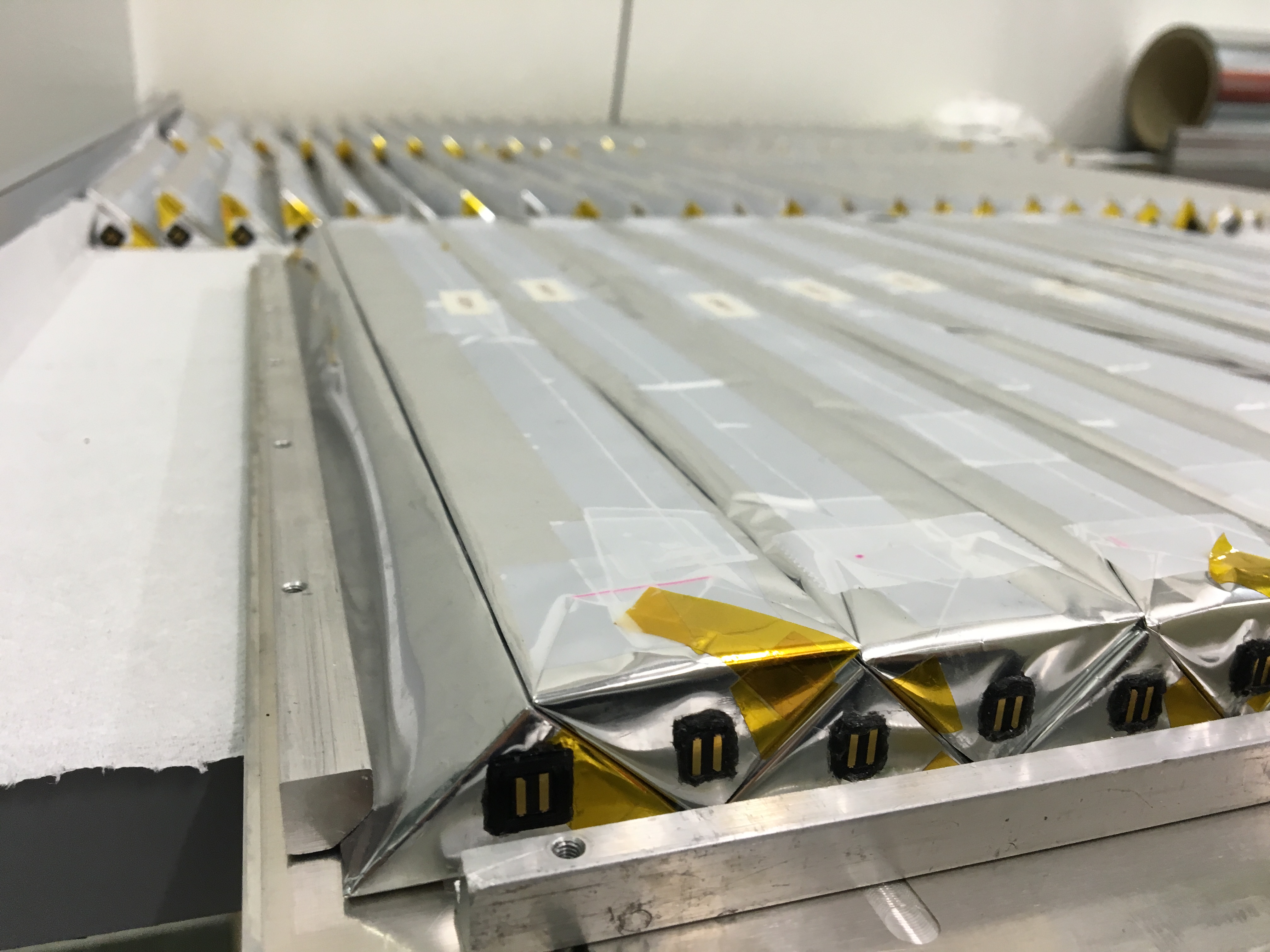}
\qquad
\includegraphics[width=.43\textwidth]{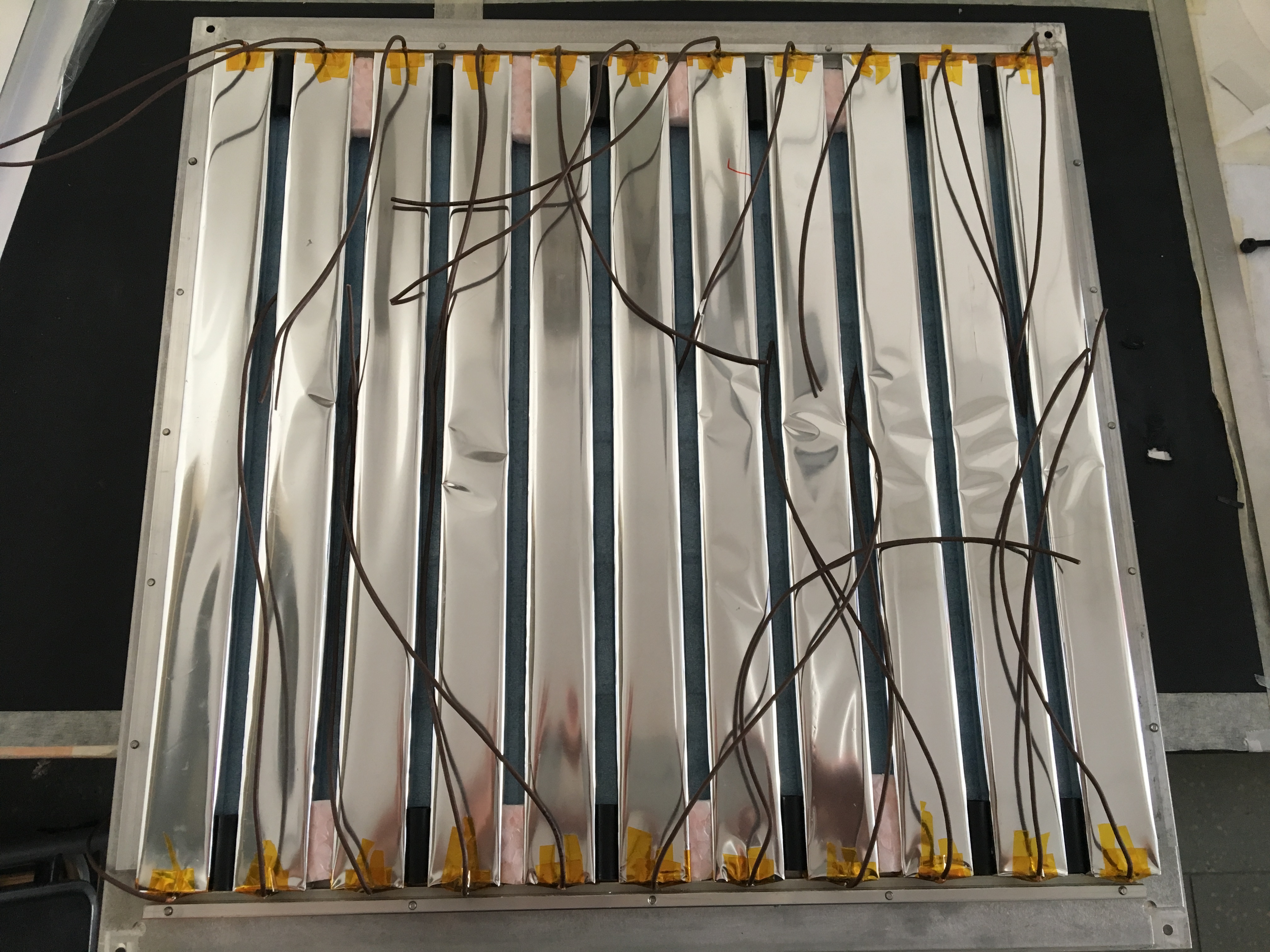}
\caption{Left: first MIMA plane built with triangular section scintillator bars. Right: one of the two MIMA planes realized with standard rectangular section scintillator bars.}
\label{fig:assembling}
\end{figure}

Each tracking plane is assembled using 21 scintillating bars, arranged as shown in Fig. \ref{fig:assembling}. The bars are contained into an aluminum frame made of two square covering plates (50\,cm $\times$ 50\,cm) separated by four small columns placed at the four corners. The mechanic's project is designed in order to minimize the total weight of the telescope but preserving the global robustness, thanks to a set of lightening millings.
To avoid unwanted movements of the bars along the two directions of the plane, that could degrade the spatial resolution of the plane, the bars are kept in position by four aluminum crossbeams. The distance between the center of the first and the last bars is fixed at 40.0\,cm with an uncertainty of 1\,mm. Thin layers of a soft plastic material, approximately 2\,mm thick, have been placed between the scintillator bars and the upper and lower covering aluminum plates, to avoid a possible damage of the Mylar foil and the eventual movement of the bars along the direction orthogonal to the plane (Z-axis). Four thin aluminum layers are fixed on the four sides of the plane to close the module and guarantee a better optical insulation of the whole plane from the outside environment. An additional layer of polyester silver tape is used to further improve the optical insulation of the planes. The upper covering aluminum plate of each plane is used to house also the front-end electronics board serving the plane itself, described with some details in section \ref{sec:easi}.
Finally the mass is approximately 7,5\, kg for each plane assembled using triangular-type bars and a bit less for those assembled using the rectangular ones. In figure \ref{fig:assembling} the inner disposition of the scintillator bars in both cases is shown.
Two aluminum support plates keep the planes in fixed positions and parallel to each other. In the standard configuration the distance between the X and Y planes of a single module is fixed at 3.1\,cm and the distance between the geometrical center of the outer X (or Y) planes is 34\,cm. A top covering panel is designed to house the DAQ electronics board and presents some dedicated holes to allow pulling cables between it and the front-end boards. Simple light aluminum plates cover all the other faces, finally giving the tracker a cubic external shape. A dedicated altazimutal mounting has been designed and produced for the MIMA cubic detector to allow defining the telescope's pointing direction by fixing the azimuth angle as desired and the zenith angle with rotations in steps of five degrees (see right side of figure \ref{fig:mima01}). The full weight of the detector, including the rotating supports, is about 60\,kg.

\section{The photosensor}
\label{sec:sipm}

\begin{figure}[thb]
\centering
\includegraphics[height=0.26\linewidth]{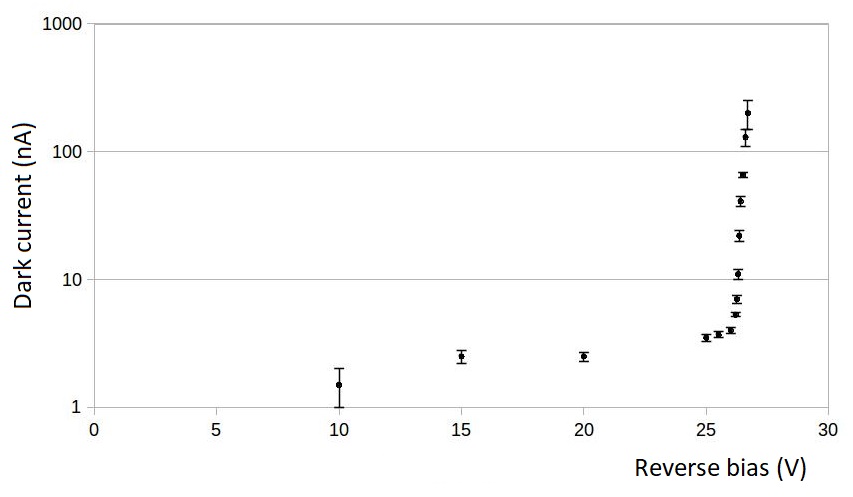}
\qquad
\includegraphics[height=0.26\textwidth]{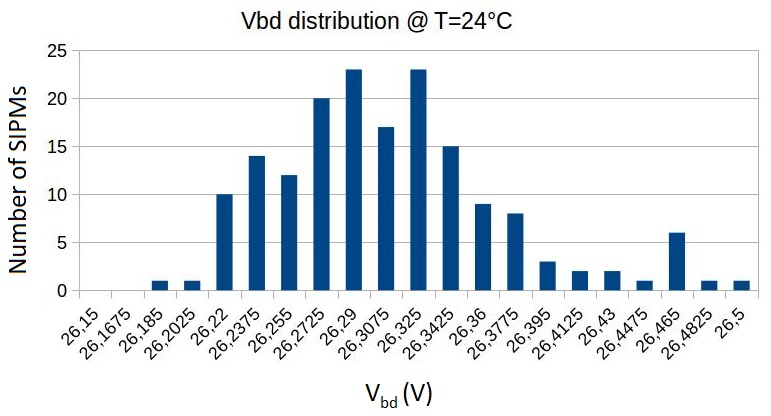}
\caption{Left: typical IV plot measured for a SiPM AdvanSiD model ASD-NUV4S-P. Right: distribution of breakdown voltage values measured for the whole set of SiPM sensors used for MIMA.}
\label{fig:sipm}
\end{figure}

Silicon photomultipliers are state of the art photosensors widely used for light detection in many on-going experiments in the field of high energy physics. This kind of devices have been chosen for the MIMA scintillator planes because of their fast response, low electric power consumption, low-voltage operation, robustness, compactness and cost-effectiveness. They consist of an array of avalanche photodiode (APD) cells working in self-quenching Geiger mode, highly miniaturized and integrated. They are characterized by a higher light detection efficiency (up to 40$\%$\,--\,60$\%$ at the wavelength of the maximum of efficiency) and a gain comparable to traditional PMTs (of the order of 10$^6$). Although each APD works in a digital way, giving a signal of fixed intensity when hit by one or more photons, the SIPM is a linear device since all the APD cells are readout in parallel, obtaining an output signal which is the sum of the signals of all the APD cells hit by photons. This property holds if the incident photon density is much lower then the cell density otherwise saturation effects become important.
SiPM sensors are capable of single photon detection sensitivity, are very fast (rise time of the order of $\simeq$1\,ns) and have extremely good time resolution.
\begin{table}[th]
  \centering
  \caption{\label{tab:sipm} ASD-NUV4S-P photodetector characteristics at T\,=\, 20$\degree$ C .}
  \smallskip
  \begin{tabular}{|lr|lr|}
  \hline
 Detection area & 4 $\times$ 4 mm$^2$ & Peak Sensitivity Wavelength & 420\,nm\\
  Dimension cell & 40\,$\times$\,40 $\mu$m$^2$ & Breakdown voltage & 26\,mV\\
  Number of cells & 9340 &Gain & 3.6\,$\cdot$\,10$^6$\\
  Quenching resistance & 800\,k$\Omega$ & Breakdown Voltage Temp. Coeff. & 26\,mV/$\degree$C\\
  Fill factor & 60\,$\%$ &Dark Count Rate ($\Delta$V$_{OV}$ = 2\,V)& $\le$ 50 kHz/mm$^2$ \\
  PDE & 43\,$\%$ & Refractive index of epoxy resin & 1.51\\
  \hline
  \end{tabular}
\end{table}
The SiPM selected for the MIMA scintillators are ASD-NUV4S-P type produced by AdvanSiD \cite{advansid} (left side of figure \ref{fig:sipm}). This model has a 4\,$\times$\,4\,mm$^2$ active area and is assembled in a Chip Scale Package (CSP) of plastic material. The active area is completely covered with a protective transparent epoxy layer where the wire bonding connecting the SiPM electrodes to the soldering pads are also embedded. This type of SiPM is appropriate for detection of Near Ultraviolet Light (NUV) since they present a peak of efficiency at 420\,nm. For this reason they result the best choice for the direct readout of the plastic scintillator material used for the MIMA hodoscope (emission peak in the violet at a wavelength of 418\,nm). The main characteristics of the ASD-NUV4S-P SiPM are listed in table \ref{tab:sipm}.
All the sensors used for MIMA were tested and an I\,--\,V characterization at room temperature was performed for each device using an automatic test station in the INFN clean room in Firenze. From the I\,--\,V curve the breakdown voltage of each SiPM was estimated. A study of the {\it dark current} versus the reverse bias intensity was performed. Left side of figure \ref{fig:sipm} shows the I\,--\,V characteristic curve measured for one device at 20\,$\degree$C. The dark current remains below 0.2\,nA for reverse bias intensities below approximately 26\,V and rises up quickly to 10\,$\div$\,100\,nA for few volts over-voltage. The breakdown voltage has been defined as the voltage for which the dark current crosses the 10\,nA level.

On the right graph of figure \ref{fig:sipm} the distribution of the measured values of the breakdown voltage $V_{bd}$ for all the SiPM used is shown. The values differ at maximum of approximately 0.2\,V, thus confirming the good production reproducibility achieved by AdvanSiD. The measurement of the $V_{bd}$ values for all the SiPM devices allow correcting the power supply voltage for the single device in such a way to equalize the gain.

\section{Front End Electronics}
\label{sec:easi}
The electronics used to read-out the photosensors is derived from the experience gained within the Mu-Ray and MURAVES projects. The read-out system is made of six independent boards, called {\it slave boards}, one for each of the six planes. In the current version of the MIMA electronics these boards are exactly the same developed for the MURAVES project \cite{MURAVES}, but a development of a new electronics is on-going to simplify the whole system and reduce further the power consumption. Currently these boards are based on the EASIROC chip \cite{easiroc}, a 32 channels analogue front end ASIC developed by Omega to readout SiPM photosensors. The six MIMA readout boards are controlled by a single master board that is used to codify and transmit the configuration instruction to the EASIROC boards, to implement the logic generating the global trigger signal for the whole system and to collect the output data flux from each board and store it temporarily before transferring it to a Raspberry $\pi$ \cite{RPI} provided of an on-board 64\,GB SDHC memory card. The MIMA hodoscope, as the Mu-Ray and MURAVES telescopes, is thought to be deployed and operated on inhospitable sites where no services are made available and the whole system has been then intended to be rugged and to have a lower power consumption (roughly 30\,W for MIMA), thus allowing to be powered also by means of a small set of photovoltaic panels. These characteristic facilitates the operation of the MIMA detector also in sites and situations where no mains electricity is available (borehole, caves, mines, archaeological or geological sites) and allows the use of a system of not excessive dimensions based on few solar panels and battery units. The operating system on-board the Raspberry computer is setup to switch on the device and start data acquisition as soon as the powering voltage is available. A network connection is therefore not required, but it might be useful to allow monitoring the status of the instrument.

\subsection{The summing circuit}
\label{sec:sumcard}

\begin{figure}
\centering
\begin{center}\begin{circuitikz}[scale=1] \draw
(1,4)  node[circle,fill,inner sep=1pt,label=above:$\textup{HV}$](hv){}
to [R=$\unit{1}{\kilo\ohm}$] (4,4)
(4,2) node[ground]{} to [C=$\unit{100}{\nano\farad}$] (4,4)
(4,4) --(5,4)--(5,2)
(9,4) to [R, l_=$\unit{47}{\ohm}$] (7,4)
(7,4) to [Do, l_=$\textup{SiPM A}$] (5,4)
(9,2) to [R, l_=$\unit{47}{\ohm}$] (7,2)
(7,2) to [Do, l_=$\textup{SiPM B}$] (5,2)
(9,2)--(9,4)--(10,4)
to [R=$\unit{1}{\kilo\ohm}$] (10,2) node[ground]{}
(10,4)--(11,4) node[circle,fill,inner sep=1pt,label=above:$\textup{Out}$](out){}
; \end{circuitikz} \end{center}
\caption{Circuit diagram of the sum card. This card is used to sum the signal outputs of the two SiPMs glued on two opposite sides of the same scintillator bar.}
\label{fig:scheda_somma_circ}
\end{figure}
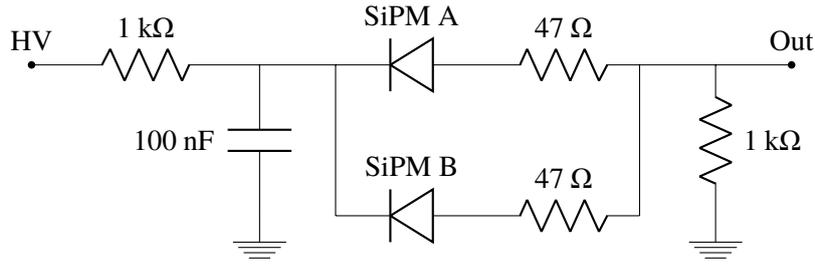

\begin{figure}[thb]
\centering
\includegraphics[width=0.8\linewidth]{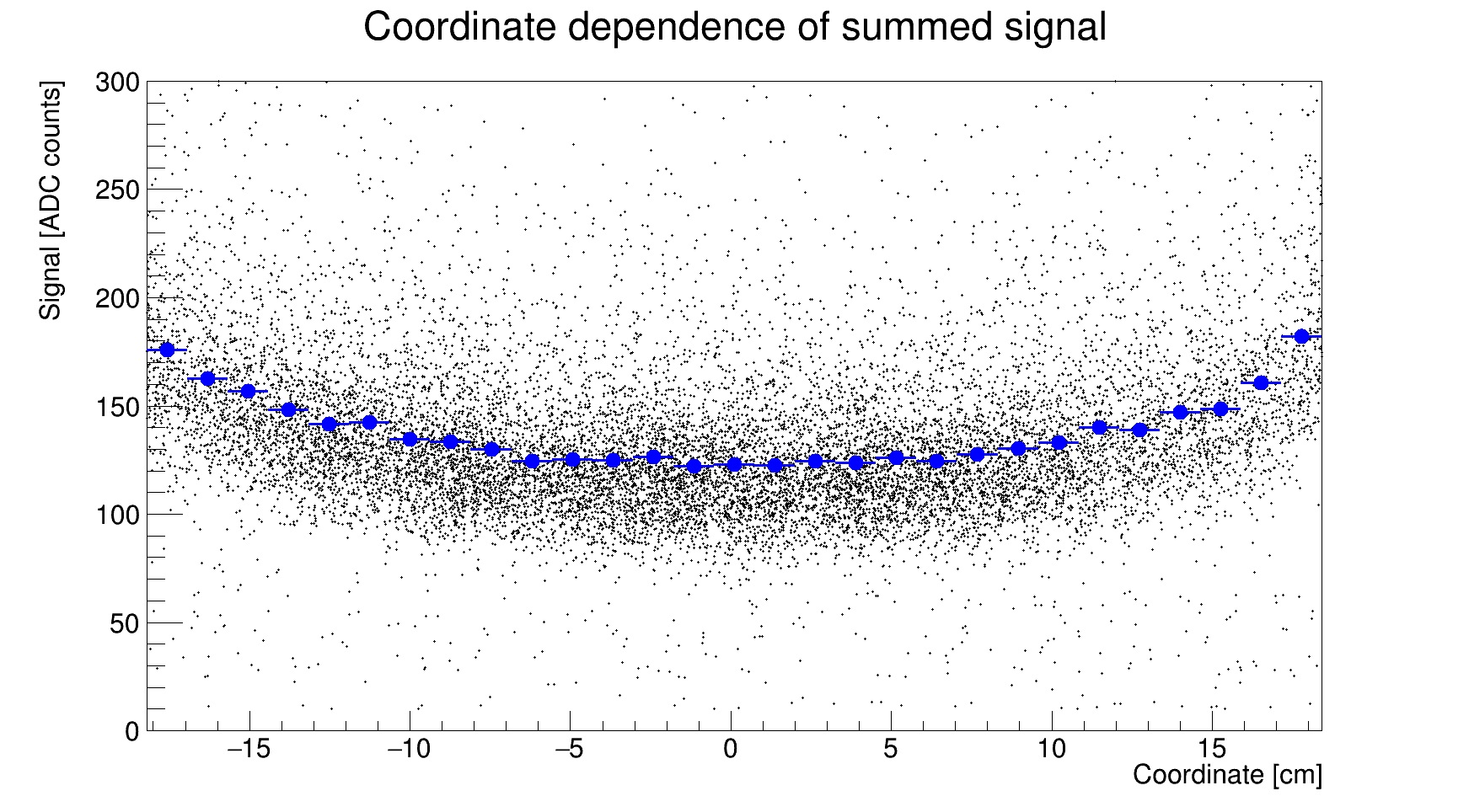}
\caption{Position dependence of the total collected signal generated by the passage of particles through two adjacent bars of a single tracking plane, along the bar longitudinal direction.}
\label{fig:signal_vs_coord}
\end{figure}

Each scintillator bar is readout by two SiPM sensors. This choice has been done to reduce the dependence of the collected signals for particles passing in different positions along the bar length. In fact the bars are quite long with respect to their transverse size, thus determining a strong dependence of the amount of light collected at each end on the particle impact point. The analog signals produced by the two photosensors, placed at the two opposite ends of the bar, are added together on a custom electronic board. The circuit diagram is shown in figure \ref{fig:scheda_somma_circ}.
In this way the fraction of light lost by one SiPM for particles passing far from it is compensated at some level by the increased amount of light collected by the other SiPM and a larger signal is associated to each scintillator bar. Figure \ref{fig:signal_vs_coord} shows the dependence of the distribution of signals released by particles on two adjacent bars of a single plane, on the reconstructed position along the bars. The average profile of the distribution is also shown. The maximum variation of the gain between the central region of the bar and the ends is less than 20\% on average. The compensation is not perfect, but allows anyway having a spatial resolution that is almost independent from the impact point.
The sum card provides also the reverse bias for the SiPMs. The bias voltage, 29\,V approximately, is kept stable by means of a low--pass filter. As shown in the diagram of the circuit in figure \ref{fig:scheda_somma_circ} the two SiPMs are connected in parallel with low value resistances and their current signals are added. The output joint line is referred to ground by a 1\,k$\ohm$ resistance, to anchor it to ground in case it is disconnected from the Front End board, and then it is sent to an input channel of the EASIROC chip in order to be processed. 

\subsection{The read--out system and the EASIROC chip}
\label{sec:boards}

\begin{figure}[thb]
\centering
\includegraphics[width=0.8\linewidth]{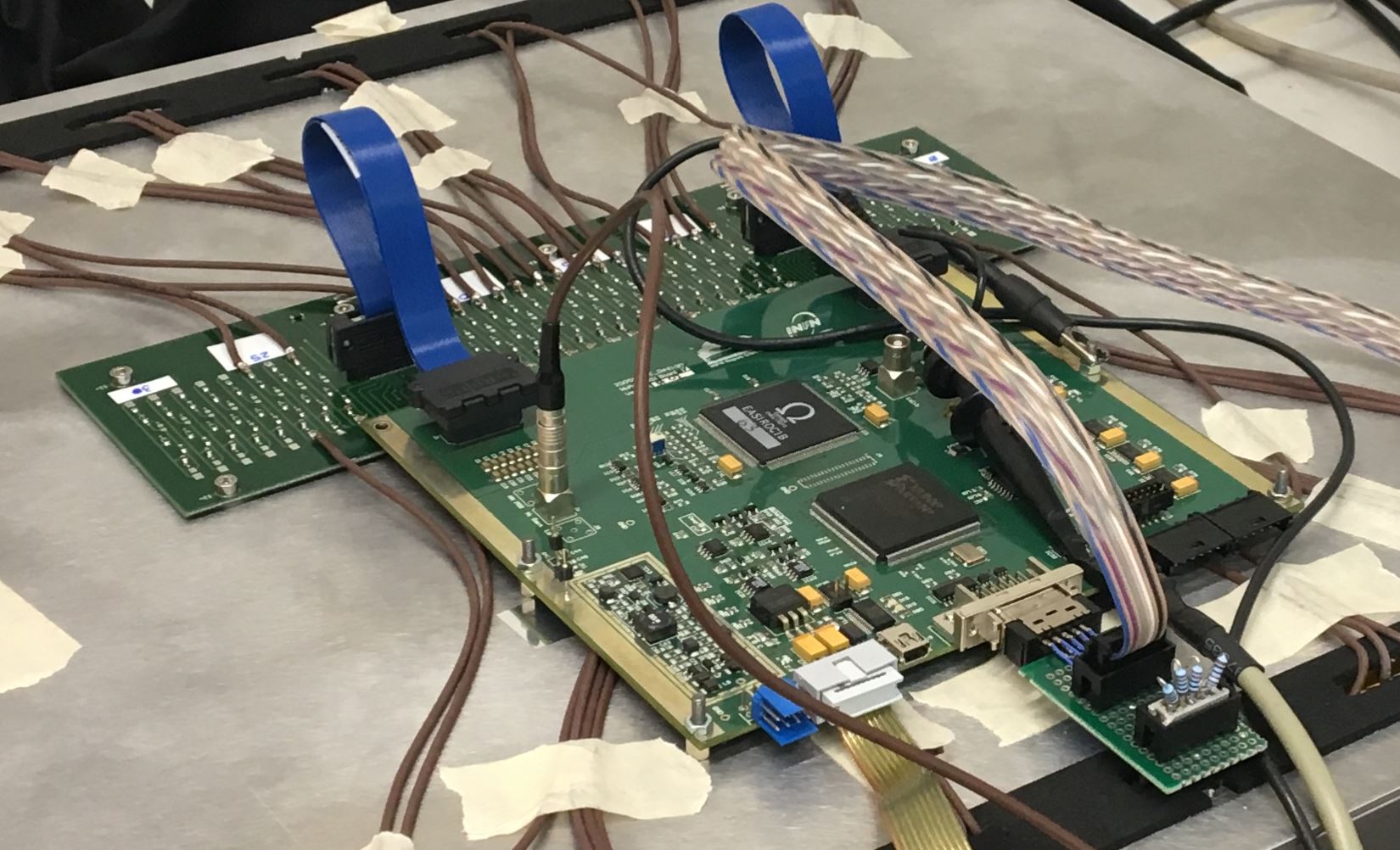}
\caption{Front End electronics mounted on a MIMA's module. On top of the module's frame the summing card (back) and the slave card (front) housing the EASIROC chip and the XILINX Spartan II FPGA are visible.} 
\label{fig:modulofinito}
\end{figure}

As shown in figure \ref{fig:modulofinito} each single plane is coupled to a sum card and to a so called {\it slave} board. Each {\it slave} board, equipped with a programmable voltage regulator (MAX1932) providing SiPM devices with a supply voltages in the range 28\,V to 75\,V, houses an EASIROC front-end chip and a XILINX Spartan III FPGA programmed to control it. The EASIROC is a 32 channels fully analogue front end ASIC ({\it Application Specific Integrated Circuit}) dedicated to readout SiPM detectors. Each input channel is provided of a 4.5\,V range 8-bit DAC (DAC8) allowing the individual SiPM gain adjustment. The incoming signal is amplified by means of two preamplifiers with different gains followed by two tuneable shapers and a track and hold circuitry. Timing capabilities are provided thanks to a trigger path integrating a fast shaper followed by a discriminator whose threshold is set by an integrated 10-bit DAC (DAC10). The output of the discriminator represents the single channel trigger signal. The logic OR of the 32 trigger signals, identified with the OR32 flag, is implemented inside the EASIROC device and defines the MIMA plane trigger signal, also defined as the ``Local Trigger'' (LT). The OR32 signals coming from all the {\it slave} boards are sent to the {\it master} board where a logic scheme can be defined to generate a ``global trigger'' (GT) signal corresponding to the events of interest. Once generated, the GT signal is then sent back to each {\it slave} board to begin reading out the data buffer produced by the active channels, whose map is defined during the chip configuration phase. Only 21 channels over the 32 available are required for a single tracking plane of MIMA, being the plane made of 21 scintillator bars. Each {\it slave} board provides therefore the necessary DAQ channels and 11 spare channels that can be used in case of any damage on the other channels, but can also be exploited to read out SiPM devices eventually connected to additional detector's parts under test. The MIMA front-end electronics is thus composed of six {\it slave} cards in total, for a total of 192 available input channels (126 of which used by default), controlled by a single {\it master} board. The power consumption is lower than 5\,mW per readout channel and additionally unused features of the EASIROC chip can be disabled to further reduce it. In case of MIMA only 21 channel are therefore usually powered ON for each EASIROC, of the 32 available.

The Master board is equipped with a credit-card-sized computer, the Raspberry Pi \cite{RPI}, housing a 64\,GB SDHC memory card where a Raspbian operating system is installed. The initial configuration of the {\it slave} boards pass from this system to the master board, that finally transmits via a serial communication protocol a configuration file with the settings of the parameters necessary for the EASIROC (low/high gain, DAC8 and DAC10, ON/OFF channel map, calibration etc.).

Since the Raspberry PI is not a real time processing device, in order to reduce the dead time between two successive events thus allowing for a faster data acquisition, the DAQ software running on the Raspberry Pi device reads data from the master board and stores them temporarily inside the on--board RAM modules. When the acquisition of the requested number of events ends, the software creates a separated thread that manages writing the data to the SDHC card. A DAQ loop is setup in such a way that the acquisition is automatically restarted. Details about the low level protocol for the communication from the Raspberry Pi and the master/slaves boards can be found in ref.~\cite{gigi}.

\section{Telescope characterization}
\label{sec:characterization}

After the construction the MIMA muon telescope was kept in operation in the INFN Firenze muon radiography laboratory for some months. Cosmic ray data sets were collected for a full characterization of the detector. Different test measurements have also been performed to explore its features. In this section we report on the study of the trigger efficiency, of the spatial and angular resolution and on the implementation of the time--of--flight capability of this system.

\subsection{Trigger system and detector efficiency}
\label{sec:trigger}

As already mentioned in section \ref{sec:boards}, the MIMA Global Trigger (GT) is produced on particular combinations of the six Local Trigger (LT) signals, that are produced inside the EASIROC chips housed on the six {\it slave} board used to readout the SiPM sensors on the six tracking planes. The LT are generated using a fast shaper followed by a discriminator each time at least one of the 32 SiPM photodetectors has a signal with an amplitude exceeding a pre-defined threshold (DAC10). The discriminator's threshold is set by a 10 bit register and the LT output signal is 100\,ns long. 

The six LT output lines are routed to programmable multiplexers on-board the {\it master} electronic card, where a trigger mask is implemented. In the standard configuration the two LT lines coming from the X and Y planes of a single tracking module are processed by a single multiplexer and their AND is requested to have an active logic output. The logic output signals corresponding to the three tracking modules of the telescope are sent then as inputs to a fourth multiplexer which is programmed to generate the final GT as the logic AND of any couple of modules inside the telescope. The configuration mask of the multiplexer system determines therefore the trigger logic of the whole detector. The global trigger thus generated is sent to all the {\it slave} boards through dedicated lines to start the data acquisition.

The choice of three modules telescope's configuration (six tracking planes, i.e. three for each view X or Y) allows the identification of the random coincidences due to the passage of different particles inside the detector's acceptance. This identification is made by imposing a cut on the goodness of the alignment of the reconstructed hits on the planes.

Dedicated measurement were performed with the aim of determining the efficiency of the planes and of the full telescope, using the trigger mask discussed in this section. Recently the trigger efficiency was studied during a measurement performed inside a mine, with the MIMA telescope positioned about 30\,$\div$40\,m underground with its pointing axis aligned along the vertical direction (i.e. with the tracking planes in the horizontal position). Data were collected for 60 days, measuring a muon trigger rate of about 0.5\,Hz.

The measured trigger efficiency for the coincidence of six planes is greater than 99\%, with single planes reaching 99.9\% of efficiency.  The efficiencies have been calculated on the base of 18600 reconstructed tracks after defining a confidence area for each plane. In particular it was required that the hits were not located within 2\,cm from the external edges of each plane, thus avoiding that events produced by muons passing in the ``peripheral'' regions of the detector and not really crossing all the six tracking planes could be accounted for as detector's inefficiencies. In table~\ref{tab:effy} the measured efficiencies for each of the six planes are reported.
\begin{table}[htbp]
\centering
\caption{\label{tab:effy} Efficiency of the single planes calculated on the basis of 186000 reconstructed tracks.}
\smallskip
\begin{tabular}{|c|cc|}
\hline
plane & efficiency & error\\
num. & in $\%$ & in $\%$ \\
\hline
0& 99.973 & 0.004\\
1 & 99.985 & 0.003\\
2& 99.849 & 0.009\\
3 & 99.989 &  0.002\\
4& 99.974 &0.004 \\
5 & 99.802 & 0.010\\
\hline
\end{tabular}
\end{table}

\subsection{Spatial and angular resolutions}
\label{sec:res}

\begin{figure}[thb]
\centering
\includegraphics[width=0.8\linewidth]{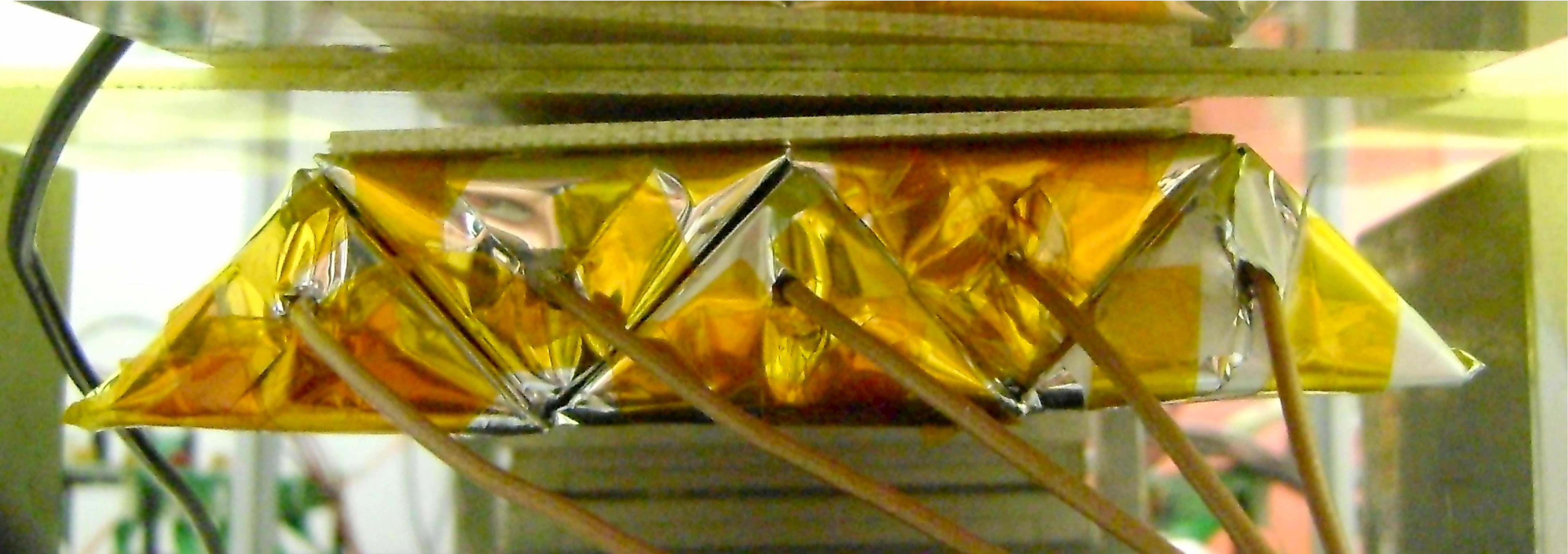}
\caption{The five scintillator bars with triangular section used for the study of the spatial resolution achievable with the MIMA tracking planes.} 
\label{fig:diletta_barre}
\end{figure}

\begin{figure}[thbp]
\centering
\includegraphics[width=0.8\linewidth]{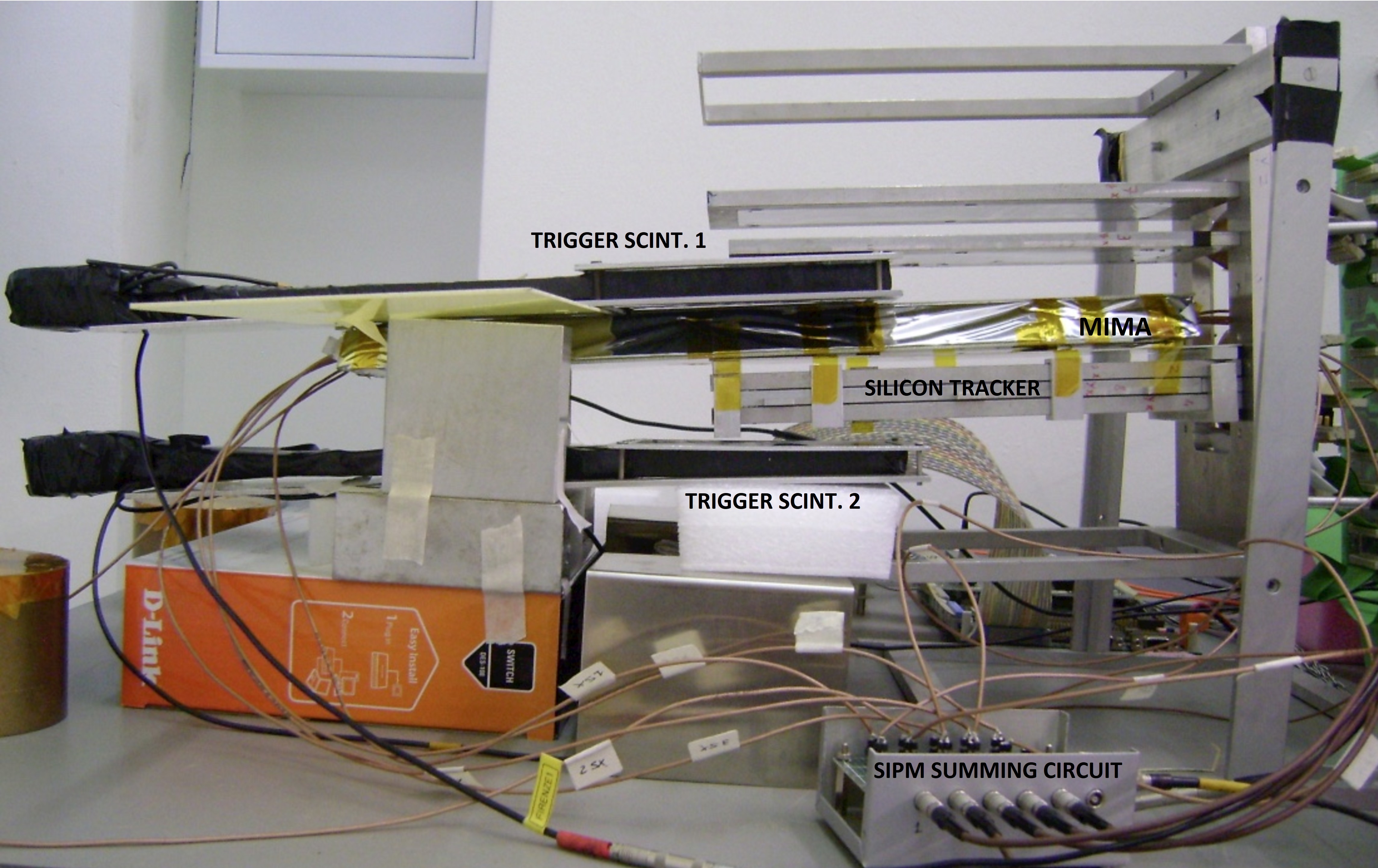}\\
\includegraphics[width=0.5\linewidth]{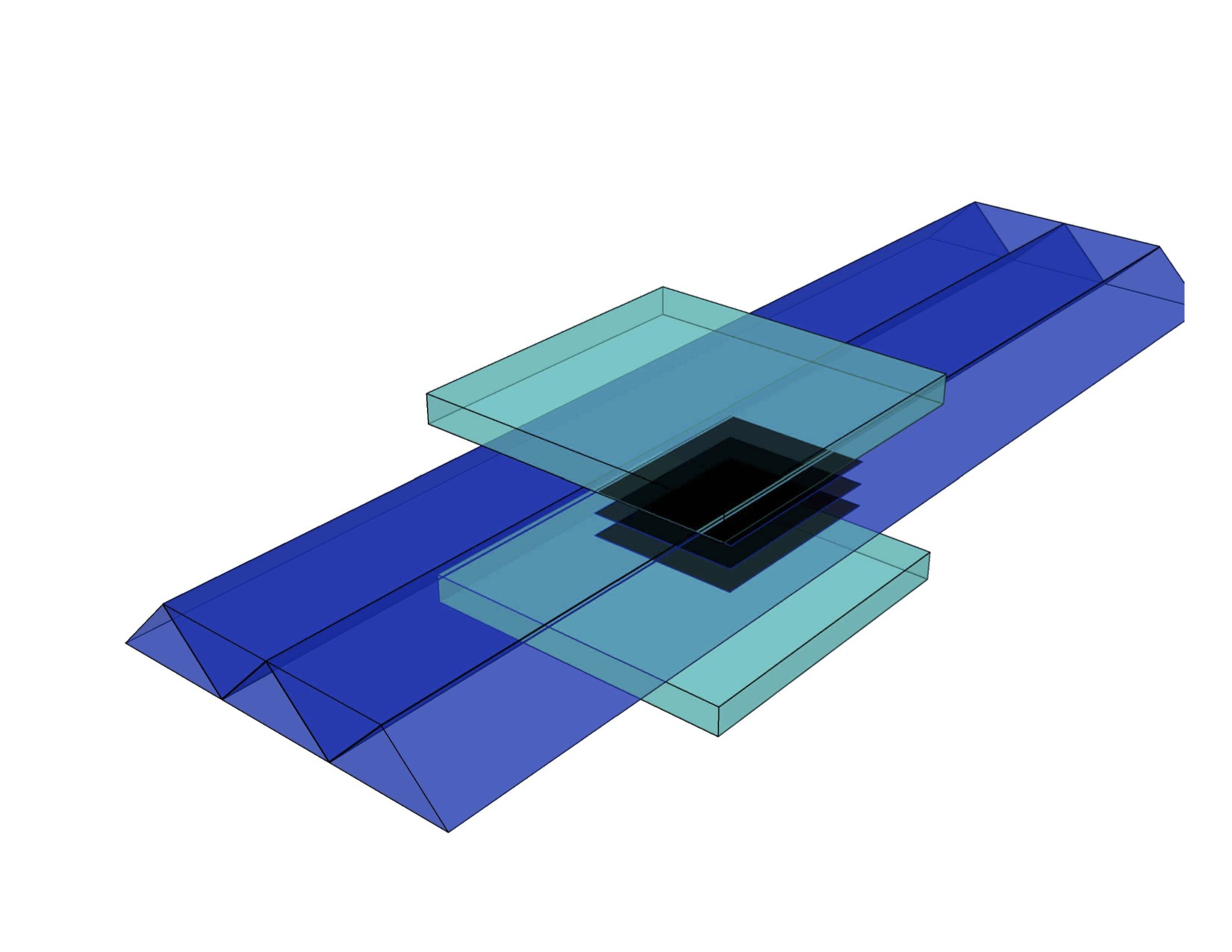}
\caption{Top: photo of the complete setup used for the study of the spatial resolution achievable with the MIMA tracking planes; the main components are indicated. Bottom: a sketch reporting the relative positions of the three types of detector used for the measurement, trigger scintillator (cyan), MIMA bars (blue) and silicon sensors (black).} 
\label{fig:diletta_setup}
\end{figure}

The spatial resolution achievable with the MIMA tracking system has been measured \cite{Diletta} using scintillator bars with triangular section (see section \ref{sec:tracking_plane}). This type of bars are currently used to build the fist two and the last two of the six tracking planes, i.e. the outer two X-Y tracking modules. As explained in section \ref{sec:tracking_plane:geom}, the triangular shape of these bars allows measuring particle's impact point as the weighted average of the charges collected by two (or more) adjacent bars, thus improving the performance achievable with a simple digital algorithm. In work \cite{Diletta} a measurement of the MIMA spatial resolution has been obtained after assembling a MIMA prototype consisting of five triangular bars reproducing the geometry of a partial MIMA tracking plane, as shown in figure \ref{fig:diletta_barre}.
The spatial resolution has been determined comparing the impact point coordinates of cosmic ray muons reconstructed using the MIMA prototype, with the same coordinates reconstructed by means of a precise auxiliary and completely independent tracking system. The auxiliary tracker was assembled using five silicon microstrip tracking modules taken from the ADAMO \cite{ADAMO} magnetic spectrometer, the same type used for the PAMELA \cite{PAMELA} satellite experiment. The complete setup is shown in figure \ref{fig:diletta_setup}.
A simplified alignment procedure allowed to determine their relative displacements, without considering the rotation angles, obtaining a spatial resolution of the order of 0.1\,mm both on the X and Y coordinates. Muon events were triggered as coincidences of the signals produced by two plastic scintillators coupled to standard PMTs. The complete data readout (MIMA+ADAMO) was implemented using a standard VME system. The two SiPM signals of each scintillator bar were summed using a handmade electronic circuit and the output lines sent to the input channels of a CAEN v792 QDC module. The silicon microstrip data were collected using a custom VME DAQ board in use within the ADAMO experiment. Only events having produced a full track in the silicon tracking system were selected to allow the comparison with MIMA.

\begin{figure}[thbp]
\centering
\includegraphics[width=0.7\linewidth]{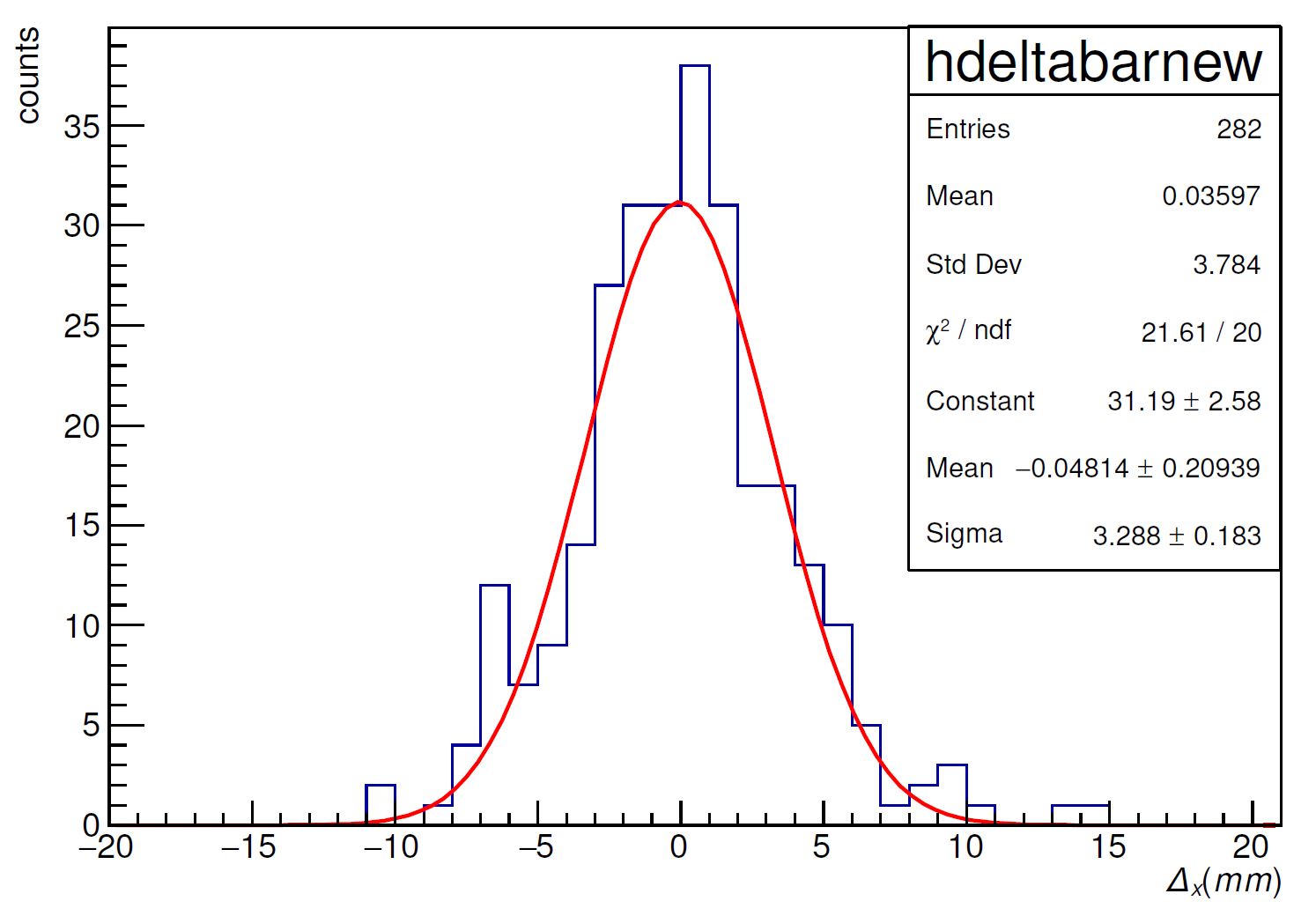}
\caption{Distribution of residuals on the measurement of muon impact points on a prototype of a MIMA tracking plane, reconstructed using the prototype itself and an independent ancillary tracking system made with silicon microstrip sensors.} 
\label{fig:diletta_spatial_resolution}
\end{figure}

In figure \ref{fig:diletta_spatial_resolution} the distribution of the difference of reconstructed values of the coordinate measured using the two systems is shown. The histogram is constructed after selecting events for which the signals coming from two adjacent bars were found over a pre-defined threshold, thus allowing for the use of a baricenter-like algorithm. In this figure a Gaussian fit of the distribution is also shown. A standard deviation $w_{x}\,\simeq\,3.3$\,mm is found, which takes into account the contributions due to the spatial resolutions of both the systems used. Because the spatial resolution of the silicon tracker is much lower than the width of the measured distribution, this is dominated by the spatial resolution of the MIMA tracking plane. We assume therefore $\sigma_{x\triangle}\,\simeq\,3.3$\,mm (where the triangle in the subscript recalls that this spatial resolution has been obtained using the center of gravity method applied to the scintillator strips with triangular section).

The spatial resolution of the MIMA tracking planes and their relative distance define the angular resolution of the muon telescope. In particular for the current configuration, ignoring the presence of the inner tracking module, currently used only for selecting straight muon tracks from random coincidences, and considering a maximum distance between the outer X or Y planes of 34\,cm, we find an angular resolution in a planar view of the order of $\sigma_{\mbox{ang}\triangle}\,\simeq\,14\,$mrad.

It has bean demonstrated \cite{Gu} that, independently on the X or Y view, the presence of a third plane located just in the middle between the outer two does not affect the angular resolution of the detector, but determines only a shift of the reconstructed tracks. For this reason there is not a major advantage to use the middle impact coordinate for the fit of the particle tracks and we are currently using it to check the quality of the reconstructed tracks. Due to the use of bars with square section instead of triangular section as for the outer planes, in the current detector's setup we expect a slightly worse spatial resolution for the central planes. After accumulating a large number of tracks with the detector axis aligned along the vertical direction, we have used events with well aligned hit points on all planes to evaluate the spatial resolution of the inner planes. Subtracting the contributions due to the outer planes, we found a value $\sigma_{\mbox{ang}\triangle}\,\simeq\,3.7\,$mm, that is not extremely worse than that found for bars with triangular section.

\subsection{Test of the Time Of Flight circuitry}
\label{sec:TOF}

The achievement of precise time-of-flight (TOF) capabilities is a crucial development in order to reduce the physical backgrounds that limits muographic measurements, especially for measurements in which the detector is pointed in a direction close to the horizontal. When the detector is pointed in the vertical direction this information is not so relevant considering that the muon flux coming from the underground is negligible. For the MURAVES measurements the detector is pointed towards the volcano almost horizontally and the flux of muons that cross the volcanic cone and arrive on the detector is very low. It is therefore necessary to identify the background due to the particles coming from the same direction in the opposite sense: these may have crossed a small part of the Earth's surface or they may have been deflected by multiple scattering from the ground behind the detector. Also for MIMA measurements for embankments monitoring it can be useful to measure the TOF.

The TOF measurement system is designed to have a reduced energy consumption (it works with a relatively low clock frequency equal to 250 MHz) and a temporal resolution of the order of 100\,ps. These requirements are satisfied using an expansion circuit for each slave card: when a Local Trigger signal is generated a first current generator loads a capacity until the Global Trigger signal arrives. Then the capacitor charging process  stops and a second current generator is activated in order to discharge the capacity more slowly. A counter measures the duration of the discharge process and provides the value of the TDC (one for each tracker plane).

The value of each $\textrm{TDC}_{i}$ will therefore be proportional to the time that elapses between the arrival of the particle signal on the electronics
and the generation of the Global Trigger. The scale factor $M$ corresponds approximately to the clock frequency multiplied by the ratio of the currents of the two generators. The latter term is the so-called expansion factor and it is about 28.

The values of $\Delta \textrm{TDC}$ calculated for the external planes, for both views, will not depend exclusively on the real TOF, which is the time it takes the muon to completely cross the detector. In fact, we have to take into account of the different delays that occur between the particle passage  through a tracker plane
and the generation of the Local Trigger signal. So we will have to consider:
\begin{enumerate}
    \item the propagation time of the light signal from particle impact point in the scintillator to the nearest SiPM;
    \item the propagation time of the electrical signal in the lemo cable from the SiPM to the sum card.
\end{enumerate}

Taking into account of these delays with a term $\Delta t_{delay}$ we obtain a proportional relation between $\Delta \textrm{TDC}$ and TOF for each view
\begin{equation}
\label{eq:TOF}
\Delta \textrm{TDC} (\textrm{TOF}) = M (\textrm{TOF} + \Delta t_{delay}) + K.
\end{equation}

Considering a different $M$ for each plane ($M_{i}$) we get a slightly different formula. Once the $M$ and $K$ are determined the previous relationship can be reversed ($\textrm{TOF}= \textrm{TOF}(\textrm{TDC}_{meas}, M, K)$).

The TOF absolute value can be geometrically calculated dividing the distance between the external impact points by the particle velocity that we assumed to be equal to $c$. Instead the TOF sign will depend on the propagation sense of the particle. For the TOF calibration, we have made measurements with the detector pointed in an horizontal direction in order to detect both the forward ($\textrm{TOF}> 0$) and backward ($\textrm{TOF} <0$) particles. To distinguish the two classes of events, assuming that almost all the particles come from above, we have imposed a geometrical cut on the direction of the reconstructed track.

With these measurements, once the track were reconstructed for each event, it was performed a calibration: we have determined the values of M and K which minimize the function
\[
\sum\limits_{events} \Biggl( \frac{\Delta \textrm{TDC}_{meas} - \Delta \textrm{TDC}(\textrm{TOF})}{\sigma _{\Delta \textrm{TDC}_{meas}}}\Biggr)^{2}
\]
Inverting the formula \ref{eq:TOF} we get two TOF estimators, one for each view. We have therefore defined the TOF final estimator ($\textrm{TOF}_{f}$) as the average of the $\textrm{TOF}_x$ and $\textrm{TOF}_y$. Eventually we used the cut $\textrm{TOF}_{f}>0$ that allows to discriminate the incoming particle direction with an accuracy of 92\%.
\begin{figure}[t]
\centering
\includegraphics[width=0.85\linewidth]{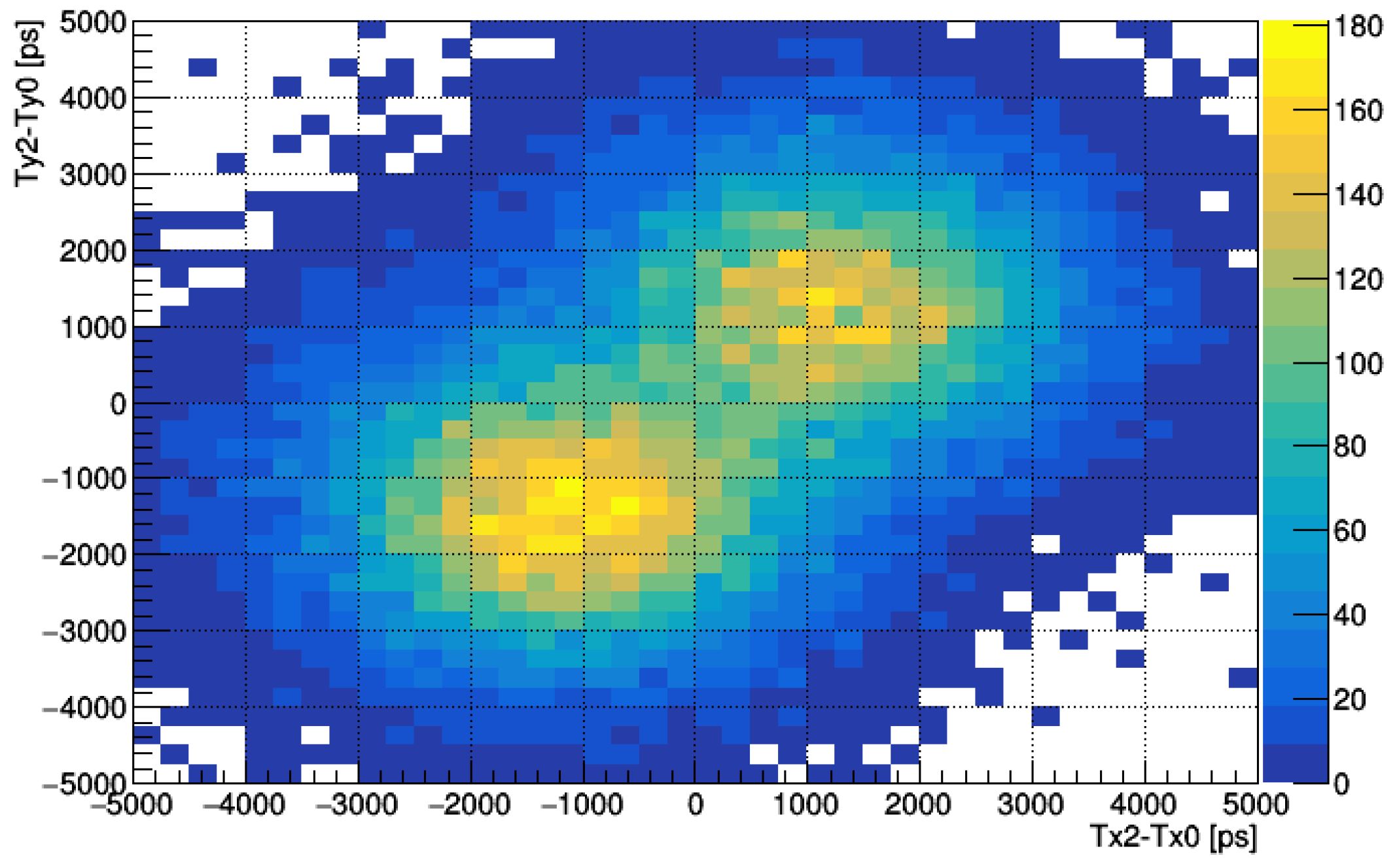}
\caption{Correlation between the Time Of Flight (TOF) values measured by MIMA exploiting separately the information coming from two X and two Y planes.} 
\label{fig:TOF_event_separation}
\end{figure}
Figure \ref{fig:TOF_event_separation} \cite{Cosimo} shows the correlation between the TOF values measured using two X and two Y planes. In this study the MIMA pointing direction was set horizontal thus geometrically accepting particles entering the detector from the first plane or from the last one. We can expect then an event distribution characterized by two peaks separated approximately of 74\,cm\,/\,30\,cm\,/ns\,$\simeq$\,2.5\,ns, in agreement with the result resported in the figure.

\section{Measurement of the free-sky vertical muon flux }
\label{sec:freesky}
In this section we report on one of the first high statistics measurement of the free sky muon flux performed with the full MIMA telescope in the current setup.

The data set selected for this analysis was taken between 14 February and 04 March 2018, for a total DAQ time of 17.3 days. 

\begin{figure}[thbp]
\centering
\includegraphics[width=0.85\linewidth]{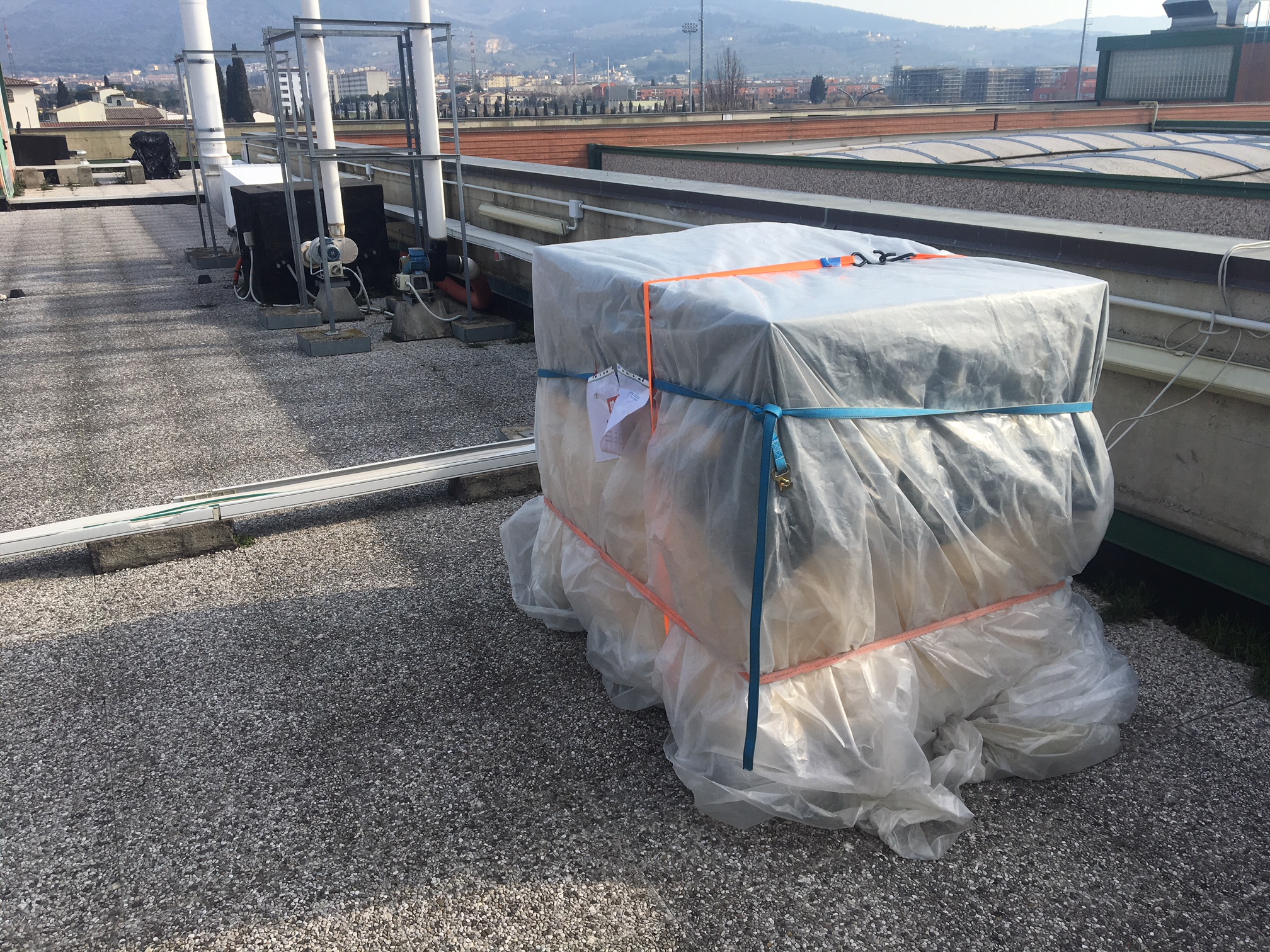}
\caption{Installation of the MIMA hodoscope on the roof of the INFN building in Sesto Fiorentino (Firenze, Italy) for a measurement of the reference free sky vertical flux.} 
\label{fig:MIMA_on_the_roof}
\end{figure}

Figure \ref{fig:MIMA_on_the_roof} shows the box containing the MIMA detector after its installation on the roof on the INFN building in Firenze. The sky was completely free within the whole acceptance of the detector. The box was well protected with a thick polyethylene film against the bad weather conditions typical of the winter time.

\begin{figure}[thbp]
\centering
\includegraphics[width=0.8\linewidth]{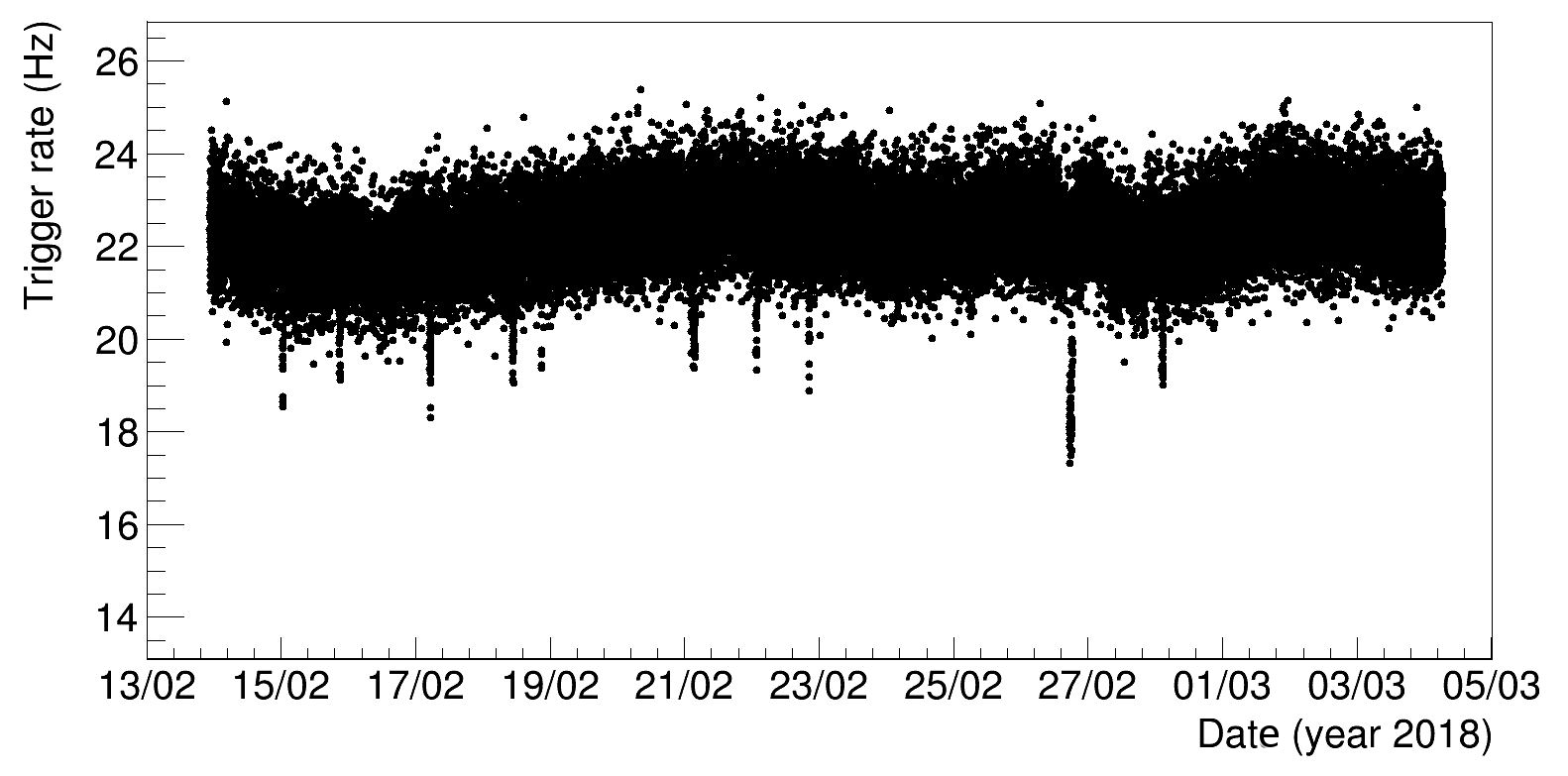}
\caption{Time variation of the total measured trigger rate during the free sky vertical flux measurement. Each point is the value of the rate evaluated over 1000 events.} 
\label{fig:DAQ_ratevst_free_sky_vertical}
\end{figure}

In figure \ref{fig:DAQ_ratevst_free_sky_vertical} the time variation of the total measured trigger rate is reported. 

\begin{figure}[thbp]
\centering
\includegraphics[width=0.8\linewidth]{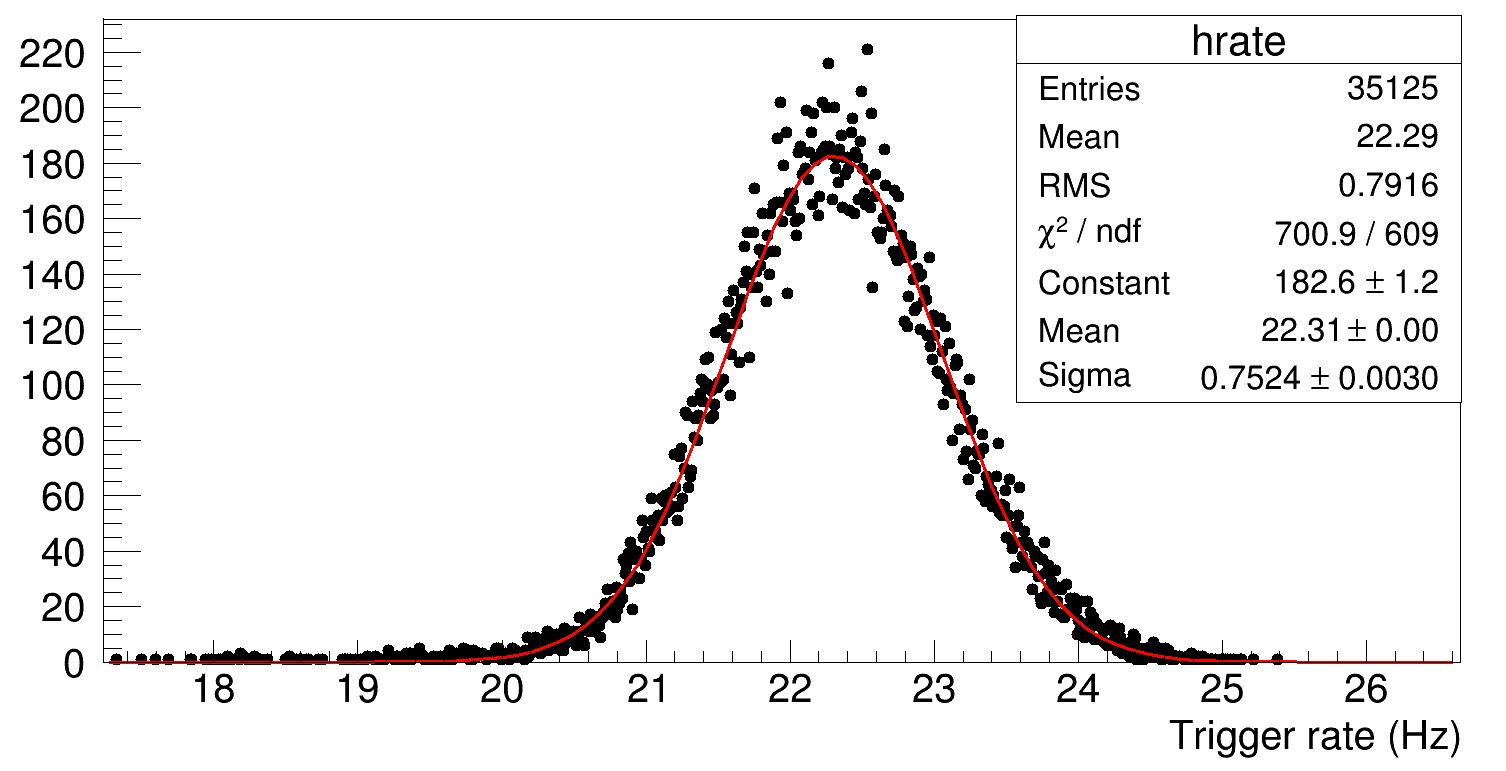}
\caption{Overall distribution of the measured event rate.} 
\label{fig:DAQ_ratedistrib_free_sky_vertical}
\end{figure}

In this plot the trigger rate is calculated for each data file, containing 1000 events in the current configuration. In this way we expect a statistical uncertainty of few percent. The total acquisition time for a single file is approximately 45\,s. This narrow binning has been chosen to identify more efficiently the single files affected by eventual problems. The distribution of the same points, projected to the ordinate axis, is shown in figure \ref{fig:DAQ_ratedistrib_free_sky_vertical}.

Some interesting features can be easily identified. The data points shown in figure \ref{fig:DAQ_ratevst_free_sky_vertical} define a band that is compatible with the statistical fluctuations on the counting. Considering the value of the measured average rate, around 22.3\,Hz, we expect a width of the distribution due to statistical fluctuations of approximately 0.7\,Hz, that is compatible with the result of the Gaussian fit of the one-dimensional distribution, shown in figure \ref{fig:DAQ_ratedistrib_free_sky_vertical}. An additional overall time modulation of the trigger rate, of the order of 2\,\%\,$\div$\,3\,\%, can be probably explained as due to the variation of the atmospheric pressure, that has an effect on the low energy muons. To be observed that this modulation cannot be due to a day/night effect, which would result in a periodicity of 24 hours. Finally looking at figure \ref{fig:DAQ_ratevst_free_sky_vertical} we find several sudden and short time intervals during which the trigger rate is systematically lower than the average value. This effect has been verified to be correlated to the activation of the procedure of data transfer through the network from the MIMA SDHC card to a remote system.

After identifying the good files, events were selected for the final data analysis as those showing particle hits aligned on all the six X-Y tracking planes, thus defining an angular acceptance of approximately $\pm\,45^{\circ}$ with respect to the MIMA axis. An extension of the acceptance is possible selecting also tracks with hits only on two X-Y modules, including tracks passing only through the central module and at least one of the outer two. In this case we use the inner module for track reconstruction and we don't have the possibility to check the quality of the track, with a slightly lower quality of track reconstruction, as previously discussed. In this way we can extend the angular acceptance up to approximately $\pm\,60^{\circ}$.

\begin{figure}[thbp]
\centering
\includegraphics[width=0.95\linewidth]{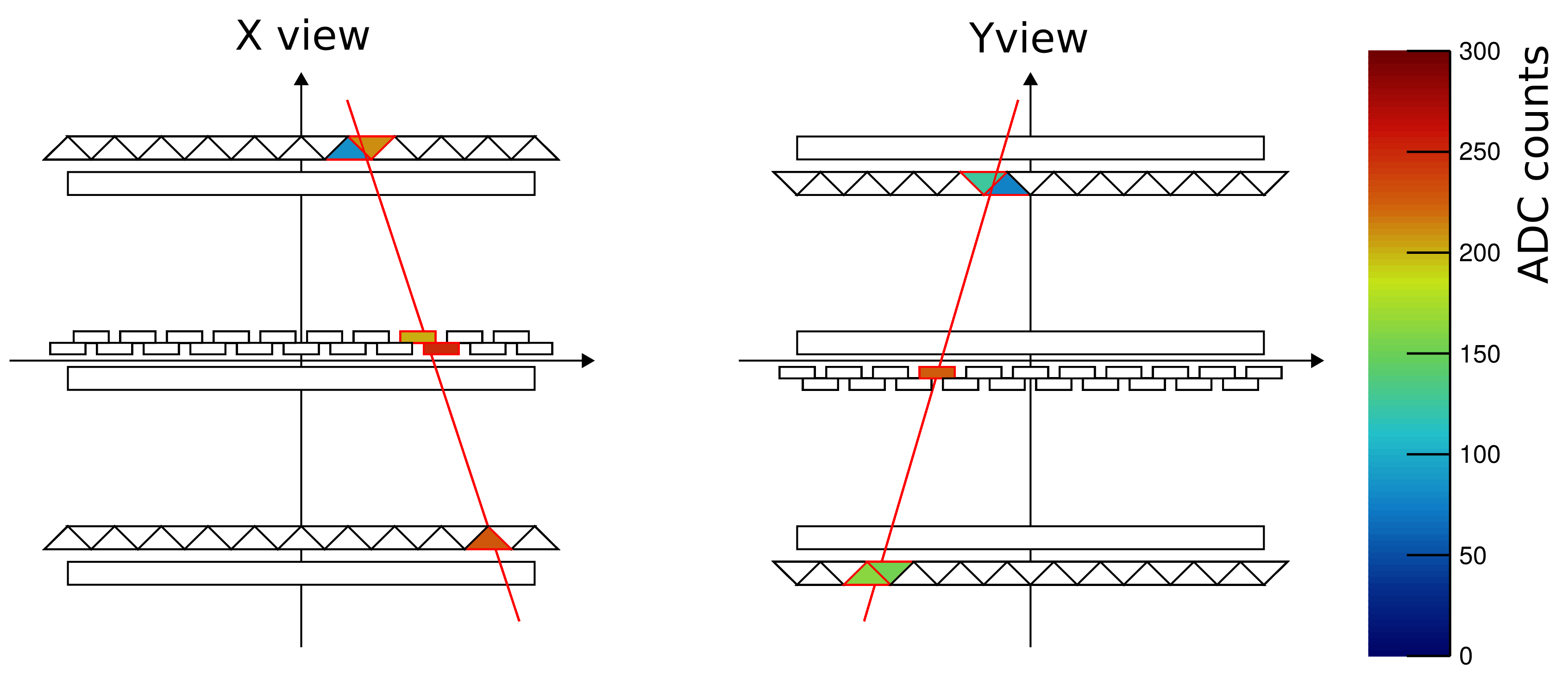}
\caption{Left side: planes measuring the X-coordinate. Right side: planes measuring the Y-coordinate. The color map shows the intensity of the signal detected on the single bars.} 
\label{fig:good_event}
\end{figure}

A typical good event measured with the MIMA muon tracker is shown in figure \ref{fig:good_event}. On the left (right) side the three X (Y) planes are shown, from the top to the bottom of the detector accordingly. The type of bars and their disposition shown in the figure correspond exactly to the current configuration. The color map on the right indicated the intensity of the signal detected on the single bars. Usually two adjacent bars are hit for each plane by muons crossing the detector, but in some case we can identify a signal over the pre-defined threshold only on isolated channels. This can be due to a lower energy release by the muon (the energy release in a thin scintillator layer is described by a Landau distribution) or to geometrical reasons (also for planes made with triangular shape bars, in case of muons crossing a bar only marginally).

\begin{figure}[thbp]
\centering
\includegraphics[width=\linewidth]{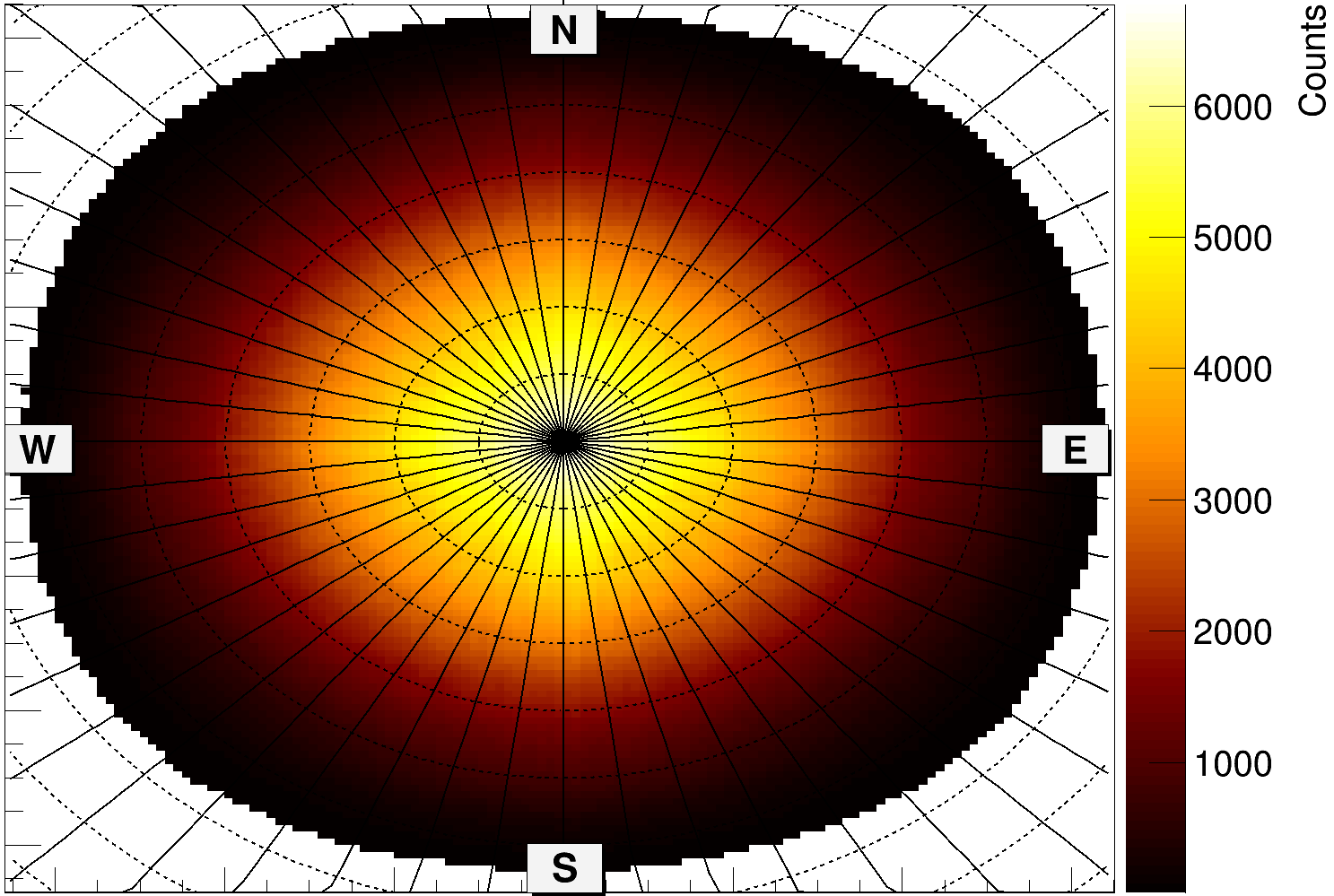}
\caption{Free sky vertical muon flux measured with the MIMA telescope. The number of identified muon events is reported in a reference frame where the distance from the center corresponds to the zenith angle and the rotation around the center corresponds to the azimuth angle. A polar grid is superimposed with $10^{\circ}$ and $15^{\circ}$ steps on the zenith and azimuth angles respectively.} 
\label{fig:free_sky_flux_polar}
\end{figure}

In figure \ref{fig:free_sky_flux_polar} the angular distribution of the whole set of selected tracks with increased acceptance is shown. Each bin correspond to a unique direction in space. The center of the plot corresponds to the vertical direction, i.e. the zenith. The distance from the center corresponds to the polar angle in a polar system with vertical polar axis. The angle defined by a rotation around the center is the azimuth angle. The incoming muon geographical reference is shown by the four labels in the outer part of the graph. 

This free sky measurement is currently used for a comparison with data taken inside the Temperino mine near Campiglia Marittima (Livorno, Italy). This mine was originally excavated and exploited by the Etruscans more than two thousands years ago and it was later exploited until the last century by different companies. More recently it was converted into a touristic site, given its importance from an archaeological point of view. Measurements with MIMA are on-going to demonstrate the capabilities of the muon absorption radiography technique in reconstructing the density profile of the inspected volumes and detecting empty cavities or dense material deposits. The results of thic measurement campaign will be published soon in a separate work. Details on the data analysis technique are not argument of this paper. Some description can be found in refs.\cite{Gu,echia}.

\section{Conclusions}
MIMA is a rugged low power state-of-the-art tracking detector designed for multidisciplinary applications of muon radiography. The MIMA's tracking planes, whose design is inspired to the Mu-Ray and MURAVES experiments, have a spatial resolution of approximately 3$\div$4\,mm in the muon impact point reconstruction. The detector's size is 50\,$\times$\,50\,$\times$\,50\,$\times$\,cm$^3$, defining a geometrical factor of 1000\,cm$^2$sr and allowing an angular resolution in track's direction around 10\,mrad. The detector can be installed on a dedicate altazimutal mounting to define precisely its measurement direction. The total weight is around 60\,kg and the power consumption of only 30\,W. All these characteristics makes MIMA suitable for installation in inhospitable environments.

The MIMA apparatus has been extensively tested in laboratory and installed in the last year in very different sites of interest for studies in the archaeological, geological and environment protection fields. In case the mains supply and the network are not available, the whole system has already been operated successfully powered by means of a standard 100\,Ah 12\,V battery connected to a 3\,m$^2$ photovoltaic panel system. In this case a modem with 3G network connection has been installed as well for remote control.

The MIMA project, with its use of low cost devices and detectors and in house equipment and software development, has the potential to extend in a very significant and inexpensive way the muon radiography technique in a multidisciplinary perspective to Archaeology, Geology, Environment Protection and other fields.

\acknowledgments
Our thanks go to our technicians involved in the design and construction of the MIMA detector, in particular to Enrico Scarlini, Massimo Falorsi, Alberto Catelani and Nicola Pasqualetti of the University of Florence, to Marco Montecchi of INFN Firenze and to Vincenzo Masone of INFN Naples.

\end{document}